# Condensate droplet roaming on nanostructured superhydrophobic surfaces

Cheuk Wing Edmond Lam[1], Kartik Regulagadda[1], Matteo Donati[1], Abinash Tripathy[1], Gopal Chandra Pal[2], Chander Shekhar Sharma[2], Athanasios Milionis[1], and Dimos Poulikakos[1, *]

[1] Laboratory of Thermodynamics in Emerging Technologies, Department of Mechanical and Process Engineering, ETH Zurich, Sonneggstrasse 3, 8092 Zurich, Switzerland

[2] Thermofluidics Research Laboratory, Department of Mechanical Engineering, Indian Institute of Technology Ropar, Rupnagar, Punjab, 140001 India

* Corresponding author

Prof. Dr. Dimos Poulikakos

Email: dpoulikakos@ethz.ch

Phone: +41 44 632 27 38

Fax: +41 44 632 11 76

**Abstract**

Jumping of coalescing condensate droplets from superhydrophobic surfaces is an interesting phenomenon which yields marked heat transfer enhancement over the more explored gravity-driven droplet removal mode in surface condensation, a phase change process of central interest to applications ranging from energy to water harvesting. However, when condensate microdroplets coalesce, they can also spontaneously propel themselves omnidirectionally on the surface independent of gravity and grow by feeding from droplets they sweep along the way. Here we observe and explain the physics behind this phenomenon of roaming of coalescing condensate microdroplets on solely nanostructured superhydrophobic surfaces, where the microdroplets are orders of magnitude larger than the underlaying surface nanotexture. We quantify and show that it is the inherent asymmetries in droplet adhesion during condensation, arising from the stochastic nature of nucleation within the nanostructures, that generates the tangential momentum driving the roaming motion. Subsequent dewetting during this conversion initiates a vivid roaming and successive coalescence process, preventing condensate flooding of the surface, and enhancing surface renewal. Finally, we show that the more efficient conversion process of roaming from excess surface energy to kinetic energy results in significantly improved heat transfer efficiency over condensate droplet jumping, the mechanism currently understood as maximum.

**Introduction**

The phenomenon of water vapour condensation on a surface begins with the formation of discrete liquid nuclei, which grow into droplets that can coalesce with one another. If not removed periodically, such droplets of condensate form a thick continuous film, which hinders the removal of heat from the vapour through the surface. The pursuit for efficient heat removal has strongly motivated surface engineering research, with the central idea to minimise the residence time and amount of the liquid condensate on the cooled surface.[1–7]

On superhydrophobic surfaces, it is possible for condensate microdroplets to spontaneously depart in the direction normal to the surface, by jumping upon coalescence, converting released surface energy to kinetic energy.[8,9] Such jumping droplet departure significantly reduces the size of droplets residing on the surface, further improving heat transfer efficiency compared to conventional dropwise condensation.[10,11] On superhydrophobic surfaces with microstructures, this spontaneous motion can also be at an angle, or even tangential, instead of normal to the surface.[12–20] It has been postulated that,[12,14–16,21] as the droplets are at the same length scale as the structures, coalescence on the side walls of the microstructure cavities ensues inclined jumping in random directions. However, tangential movement of coalescing condensate microdroplets is seen on superhydrophobic surfaces with solely nanostructures,[22,23] which are orders of magnitude smaller than the droplets in concern, with remarkable effects on ensuing heat transfer augmentation. Here we explain this unexplored droplet roaming coalescence mechanism, identify the conditions under which roaming occurs, and determine its significant effect on heat removal from a surface.

As in-plane roaming can span great lengths and coalesce with other condensate microdroplets along the way,[22] compared to out-of-plane jumping which is confined to a local cluster of droplets, it provides a pathway to continuously remove larger amounts of condensate.[12] Frequent roaming exposes needed underlying surface for new nucleation cycles, thus reducing droplet sizes against conventional gravity-driven dropwise condensation, and ultimately markedly improving heat transfer, even compared to jumping dropwise condensation.

**Results and discussion**

Roaming on solely nanostructured surfaces

To allow high-speed imaging of roaming, we prepare a reflective solely nanostructured superhydrophobic surface by exposing flat aluminium substrates to hot water to form boehmite nanowalls,[24–26] which are then coated with hydrophobic poly-(1H,1H,2H,2H-perfluorodecyl acrylate) (pPFDA) using initiated chemical vapour deposition (iCVD),[2,12] (**Methods**). The coating conforms to the nanowalls, and its thickness is measured to be 3.5 nm with ellipsometry (**Supplementary Information S1**). An image of the nanostructures with the pPFDA coating using scanning electron microscopy (SEM) is seen in **Figure 1a**. The advancing contact angle and contact angle hysteresis are 162.5° ± 1.8° and 1.1°, respectively (**Methods**).

All samples are tested in our condensation setup illustrated in **Figure 1b**. A transparent window separates the condensation chamber environment and the atmosphere. During experiment, saturated steam (30 mbar, 24.1 °C) continuously passes over and condenses on the cooled surface of the sample. A microscope objective in front of the window enables direct observation of microscale condensation behaviour with a high-speed camera at a resolution of 4.5 μm/pixel, and temperature sensors in the chamber allow the simultaneous measurement of heat transfer performance. See **Supplementary Information S2**.

We investigate roaming motion on solely nanostructured superhydrophobic surfaces and avoid the presence of microstructures which can alter the motion of coalescing condensate microdroplets at the similar length scale (**Supplementary Information S3**). A typical roaming event is shown in **Figure 1c**. After the first coalescence (Panel i), there is a tangential motion to the right (Panel ii). The droplet coalesces with other droplets along its way, roaming the surface (Panel iii), before coming to rest (Panel iv). The corresponding video can be found in **Video S1**. Evidently, roaming demonstrates in-plane arbitrary directionality which spans across considerable time and distance, and is distinct from localised multi-droplet coalescence,[27,28] which is confined to a specific location and occurs on shorter timescales (**Video S2**).

We observe the condensation behaviour as we increase the surface subcooling $\Delta T = T_{\text{steam}} - T_{\text{surf}}$, the difference between the steam temperature $T_{\text{steam}}$ and the surface temperature $T_{\text{surf}}$. Above a subcooling of $\approx 1.5$ K, roaming becomes frequent on the surface. We characterise roaming events from two perspectives: (1) we measure the position and time of all visible droplets which coalesce and take part in the event (participating droplets, shown as circles in **Figure 1d**); and (2) we track over time the location and shape of the travelling droplet (main droplet, trajectory and shape shown in inset of **Figure 1d**) which grows as it coalesces with and absorbs the participating droplets. Details of image processing, and individual droplet measurement and tracking can be found in **Supplementary Information S4**.

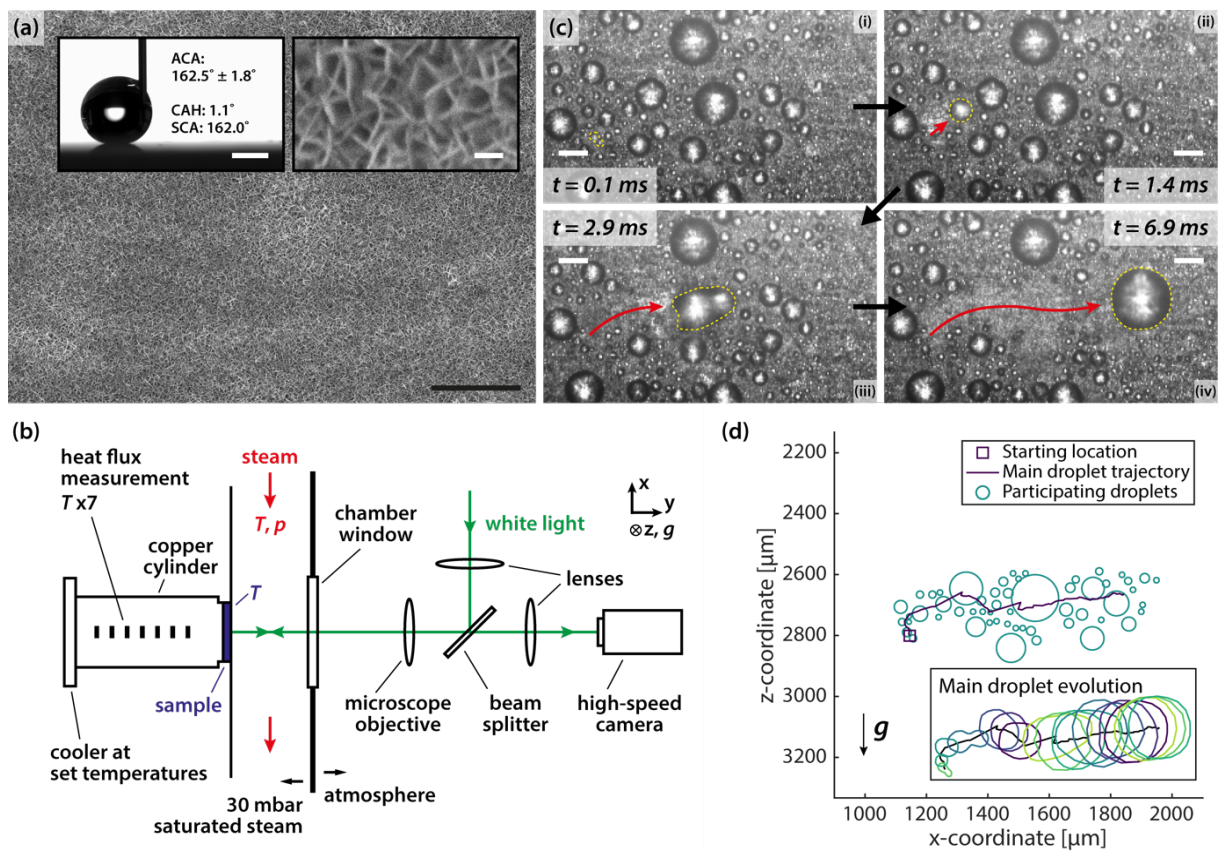

**Figure 1**: Roaming on solely nanostructured superhydrophobic surfaces. (a) SEM image of the boehmite nanowalls coated with pPFDA, a solely nanostructured superhydrophobic surface. Scale bar: 2 µm. Left inset: water droplet being deposited at 2 µL s$^{-1}$, and wettability measurements of the advancing contact angle (ACA), contact angle hysteresis (CAH), and the static contact angle (SCA). Scale bar: 1 mm. Right inset: SEM image of the nanowalls at higher magnification. Scale bar: 100 nm. (b) Schematic of the condensation and observation setup. $T$ and $p$ refer to temperature and pressure measurements respectively. Gravity $g$ is in the z-direction. (c) Roaming event during vapour condensation on boehmite nanowalls coated with pPFDA. Yellow dashed lines enclose the main droplet. Red arrow indicates the approximate trajectory of the roaming event. Also see **Video S1**. Subcooling: 2.6 K. Gravity is downwards. Scale bars: 100 µm. (d) Participating droplets distribution for the in-plane (xz) roaming event in **c**. The line represents the trajectory of the main droplet (red arrow in **c**). Inset: evolution of the shape of the main droplet. Every contour is 0.4 ms apart.

Mechanism of roaming coalescence

**Figure 2a** and **2b** display the participating droplet distributions and the corresponding trajectory of the main droplet of several roaming events, out of a total of 28 measured at a subcooling of 2.2 – 2.6 K. Roaming occurs in all in-plane directions, independent of gravity. This can be explained with the Bond number $Bo = \Delta\rho g R^2/\sigma$, where $\Delta\rho$ is the density difference of liquid water and its saturated vapour, $g$ is the gravitational acceleration, $R$ is the characteristic droplet length scale, and $\sigma$ is the surface tension of liquid water. Substituting the mean main droplet equivalent radius of 79 ± 28 μm as the length scale, Bo = 0.0009 ≪ 1, indicating that roaming is dominantly a capillary phenomenon. During our experiments, we have not seen roaming events to repeatedly occur at fixed locations on the condensing surface (i.e. the starting and terminating locations are random, see **Figure 2b**), indicating that it is not triggered by surface defects. Roaming is also found to be independent of the form of nanostructures, as we have observed its occurrence on titanium dioxide nanorods[29] and copper(II) hydroxide nanoneedles[30] as well (**Supplementary Information S5**). Occasionally, a roaming event can alter its direction as it progresses in-plane, see **Video S3**. Apart from coming to rest as in **Figure 1c**, it can also terminate by jumping (**Video S4**). Of the measured roaming events, the mean duration is 5.3 ± 3.4 ms, with a mean travelled distance of 744 ± 334 μm, on average 37× of the mean participating droplet diameter of the event.

Roaming requires significant generation of tangential momentum. **Figure 2c** displays the distance travelled over time for the main droplet of roaming events. Although the mass increase of the main droplet varies significantly for different events (**Supplementary Information S6**), it largely follows a constant initial velocity of ≈ 0.18 m s⁻¹, before slowing and diverging at approximately 3 – 5 ms as viscous dissipation sets in (compared to a viscous timescale $t_\mu = \rho R^2/\mu = 7$ ms, where $R = 79$ μm is the mean main droplet equivalent radius and $\mu$ is the dynamic viscosity of liquid water). We term this velocity of the travelling main droplet as the apparent roaming velocity.

As the main droplet gains mass and size, perturbations from further coalescence with upcoming participating droplets increasingly contribute to the low-amplitude capillary waves at the liquid-vapour interface, instead of bulk droplet motion. This is due to the increase in the number of available oscillation modes in a larger main droplet,[31,32] along with coalescence bridges becoming increasingly small compared to the traversing main droplet. As viscous effects become important, the roaming event slows and terminates. We quantify the intensity of coalescence over the course of the events by describing the shape evolution of the main droplet in **Figure 2d**. At the transition time of $\approx$ 5 ms, its circularity and Feret ratio quickly approach unity, indicating transition to a circular contour (also see inset of **Figure 1d**). At the same time, the ratio of participating droplet sizes relative to the main droplet size drops below unity (**Supplementary Information S6**).

The translational kinetic energy of roaming stems from the excess surface energy due to the reduction of liquid-vapour interfacial area upon coalescence. We compare the roaming velocity with the theoretical maximum velocity (i.e. if all excess surface energy were converted to in-plane translational kinetic energy) to quantify the efficiency of this energy conversion. However, even without tangential momentum generation, there would be a shift in the location of the main droplet after it coalesces with each participating droplet, due to the addition of mass from the participating droplet to the main droplet away from its location. Therefore, to account for this effect and extract the "real" roaming velocity that is purely the result from tangential momentum generation, we dynamically measure the increase in the distance between the main droplet and the centre of mass of the system of coalesced participating droplets. The real roaming velocity is 48 ± 14% of the theoretical maximum. For out-of-plane droplet jumping, it is $\approx$ 20%.[8,33] See **Supplementary Information S7**. Roaming better scavenges the excess surface energy of coalescence, which would otherwise be dissipated as heat, for condensate removal, improving heat transfer efficiency.

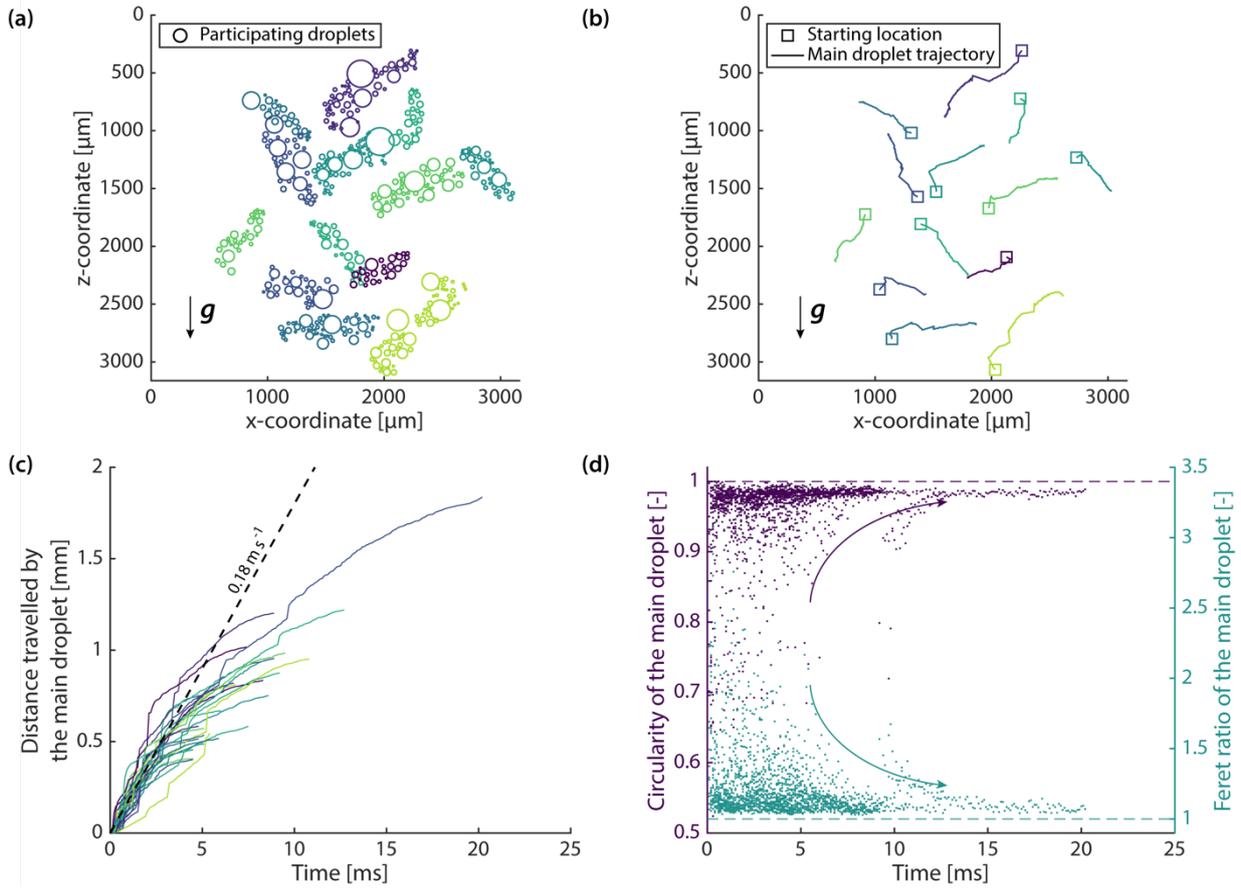

**Figure 2**: Characteristics of roaming events. (a) Participating droplet distribution of 13 selected roaming events (indicated by different colours), at different locations of the surface. Roaming events do not repeatedly occur at the same location over time. (b) Main droplet trajectories corresponding to the events shown in **a**. Squares indicate starting location of the events. All events progress in in-plane directions, independent of downward gravity. (c) Distance travelled by the main droplet for all the events. Initial roaming velocity 0.18 m s$^{-1}$. (d) Circularity ( $= 4\pi(\text{area}/\text{perimeter}^2)$ ) and Feret ratio ( $=$ maximum caliper diameter/minimum caliper diameter) of the main droplet. Both approach unity at $\approx$ 5 ms.

Roaming condensation heat transfer and the transition subcooling

The more efficient energy conversion and larger span of roaming than jumping suggest heat transfer benefits. In **Figure 3**, the heat transfer performance and the condensation behaviour at different subcooling are quantified. We first measure the heat flux $q''$ and the subcooling $\Delta T$ (**Supplementary Information S8**), and compute the heat transfer coefficient $h = q''/\Delta T$ (**Figure 3a**). A surface with the same boehmite nanowalls but without the pPFDA coating (pristine boehmite) is used as the filmwise condensation control.

Overall, the superhydrophobic boehmite surface is superior: when comparing the mean of all measurements from each surface, there is an increase in the heat transfer coefficient of over 300% from 20.1 kW m$^{-2}$ K$^{-1}$ on pristine boehmite to 82.6 kW m$^{-2}$ K$^{-1}$ on the superhydrophobic surface. However, on the superhydrophobic surface alone, there are two regimes of condensation mode, dependent on the current subcooling. At low subcooling, condensation is dominated by the jumping dropwise mode with a relatively lower heat transfer coefficient (mean = 62.7 kW m$^{-2}$ K$^{-1}$, first 3 points from the left in **Figure 3a**); but when subcooling increases, there is a transition and condensation is dominated by the roaming mode with an increased heat transfer coefficient (mean = 97.5 kW m$^{-2}$ K$^{-1}$, last 4 points from the left in **Figure 3a**). When compared to filmwise condensation at similar subcooling (mean = 25.4 kW m$^{-2}$ K$^{-1}$, first two points from the left in **Figure 3a**), jumping dropwise condensation provides a 147% increase in the heat transfer coefficient while roaming condensation provides a 284% increase. The synergistic effect of a higher heat transfer coefficient at a higher thermal driving force, i.e. subcooling, results in a 175% higher heat flux for roaming condensation than jumping dropwise condensation. The jumping-roaming transition can be seen in **Figure 3b** and **Video S5**. We quantify the transition in the top subplot of **Figure 3c**, and show that when the subcooling increases past the transition at ≈ 1.5 K, the surface area renewal rate $S'$ from roaming sharply increases. Remarkably, over 70% of the surface is renewed every second by roaming when it is the dominant mode. See **Supplementary Information S9**.

The roaming mode provides higher heat transfer efficiency than the jumping dropwise mode despite a surface with more larger droplets as seen in **Figure 3b**. In condensation, most of the heat and mass transfer is attributed to the initial droplet growth after nucleation.[34] On a surface with a distribution of various condensate droplet sizes, the majority of heat flows through the smallest droplets. When subcooling increases, (1) the diameter at which nucleation occurs, i.e. the critical nucleation diameter $d_{crit}$, decreases, and (2) the nucleation rate (number of nuclei per area per time) increases.[35] Therefore, at elevated subcooling, a renewed surface area is soon filled with a large number of small condensate droplets ideal for heat transfer. Roaming maximises the heat flux by facilitating the frequent renewal of large surface areas, enabling abundant nucleation.

The emergence of roaming when subcooling increases provides a clue about its origin. Refer to **Figure 3c**. As $d_{crit}$ is reduced with increasing subcooling and becomes smaller than the nanocavity sizes, nucleation occurs stochastically within the nanocavities. At the transition subcooling of 1.5 K, $d_{crit}$ ($\approx$ 23 nm) is below the majority of boehmite cavity sizes (**Figure 3c**). On the other hand, we find that on surfaces with much sparser nanostructures and thus larger cavities such as copper(II) hydroxide nanoneedles, the transition subcooling is notably reduced to 0.7 K (**Supplementary Information S10**). In addition, at a subcooling of 1.3 K, these sparser copper(II) hydroxide structures begin to flood as most of the nanocavities are filled with condensate. This jumping-roaming-flooding transition evidently shows that roaming is closely related to condensate filling of *some* nanocavities, producing *some* droplets on the nanostructures which are in the partial-Wenzel state.

For a cavity to be filled, two competing factors are in play. When subcooling increases, although the nucleation rate increases, the volume of each nucleus reduces. To assess the probability of cavity filling through nucleation, we define a volumetric nucleation rate as the product of the two. A substantial increase in the volumetric nucleation rate is seen around the transition subcooling at 1.5 K, indicating the increased likelihood for the cavities to be filled. The timescale to fill the nanocavities of boehmite is in turn estimated to be 0.1 – 1 ms. See **Supplementary Information S11**.

As roaming is not observed to repeatedly initiate at certain locations (**Figure 2b**), and the surface can sustain roaming condensation at steady state with no surface flooding over time, such stochastic locally wetted cavities are expected to dewet in a roaming event, similar to the case of dewetting by condensate motion previously observed[8] in the case of droplet jumping.

Such dewetting brings perspective about condensation on superhydrophobic structured surfaces. Effective nucleation rates on structured surfaces are higher due to the extra area from the roughness when the critical nucleation diameter is small enough to nucleate within the structure cavities, enabling higher heat fluxes. However, these higher nucleation rates are only sustainable when balanced with enhanced condensate removal from the cavities, so that they do not become permanently filled to conceal the extra area. Micro- and/or nanostructures which promote the ejection of droplets from within are often exploited to achieve this goal.[36–40] As jumping subsides at higher subcooling, roaming provides an alternative and regarding heat transfer, more efficient pathway to utilise the excess surface energy of coalescence for cavity dewetting and thus the sustenance of further increased nucleation rates and smaller nuclei. Lastly, the fact that some droplets are in the partial-Wenzel state also benefits heat transfer due to their larger contact area with the substrate than Cassie-state droplets.[41]

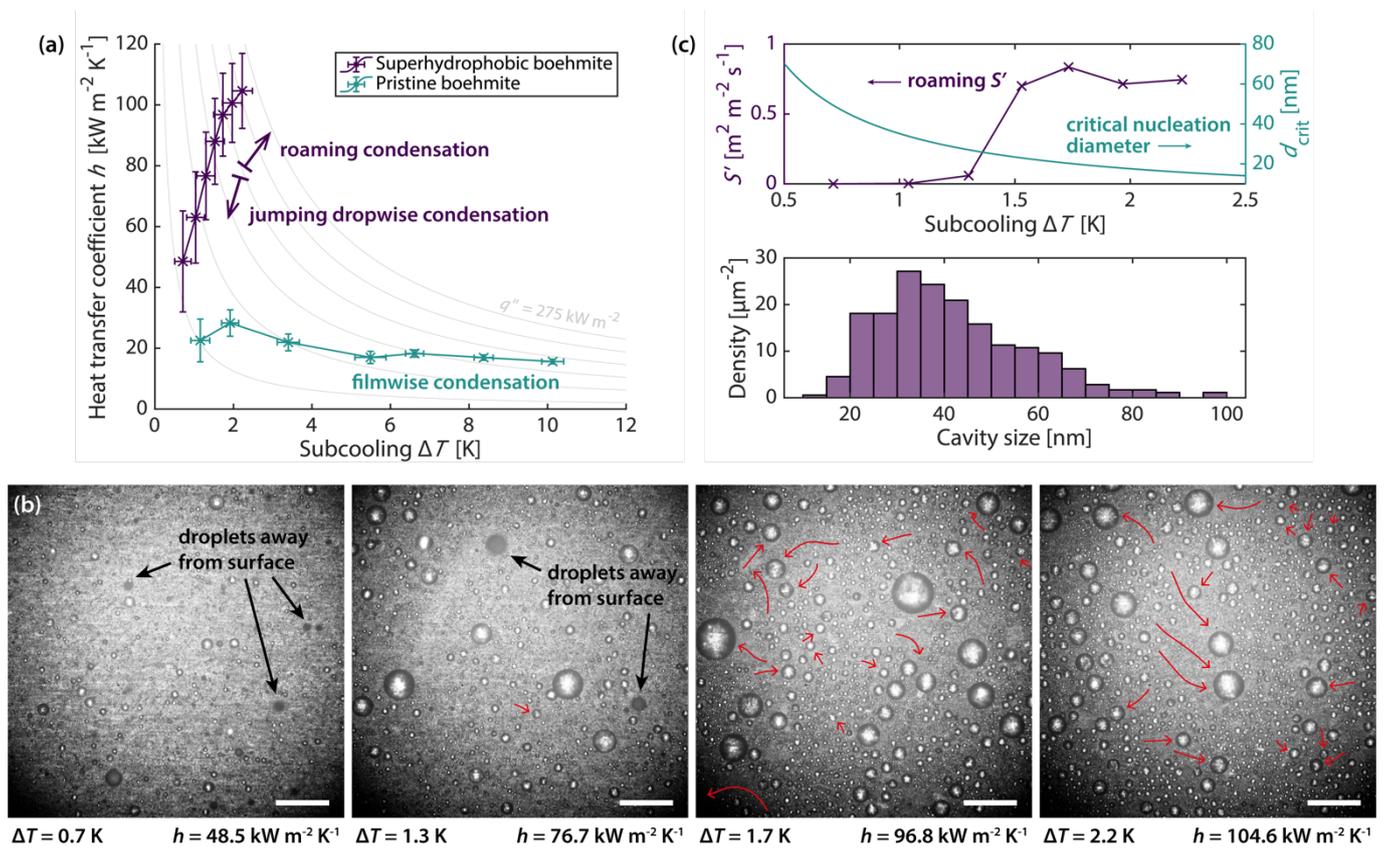

**Figure 3**: Heat transfer performance of roaming condensation. (a) Heat transfer coefficients $h$ at steady state. Lines of constant heat flux $q''$ are shown in grey, from 25 to 275 kW m$^{-2}$ at intervals of 50. For a fair test, the 7 subcooling achieved for each surface correspond to 7 identical cooler temperatures. The achievable subcooling range is smaller for a surface with better performance, as the surface temperature would closely approach the steam temperature in the same system. On the superhydrophobic surface, two modes of condensation are observed. (b) Snapshots of condensation behaviour for superhydrophobic boehmite. Transition is seen from jumping dropwise to roaming condensation. At the lowest subcooling (0.7 K), only jumping is observed and there are numerous droplets in the vapour. At 1.3 K, the number of jumped droplets in the vapour is visibly reduced, and some are seen to return to the surface. After the transition subcooling (≈ 1.5 K), condensation is dominated by roaming. Red arrows are trajectories of roaming events. See **Video S5** for the corresponding video. Scale bars: 500 μm. (c) Surface area renewal rate $S'$ from roaming (unit: m$^2$ of surface area renewed per m$^2$ of condensing surface per s of time) and critical nucleation diameter $d_{\mathrm{crit}}$ for 30 mbar saturated steam (top). As transition to roaming occurs at ≈ 1.5 K, the critical nucleation diameter lies below most nanostructure cavity sizes (bottom).

Generation of tangential momentum in roaming

The absence of roaming at the limit of low subcooling suggests that this phenomenon is exclusive to condensate droplets – gently deposited Cassie-droplets do not roam upon coalescence. Indeed, no tangential motion has been reported in the literature for deposited droplets on solely nanostructured surfaces. As some nanocavities fill, stochastic wetting at random locations across the surface promotes the concurrent presence of Cassie and (partial-)Wenzel-state condensate droplets on superhydrophobic surfaces.[41] Their different adhesion[39] produces an adhesion asymmetry of the droplets for tangential motion. Without this, there is no apparent reason for out-of-plane jumping to transition to in-plane roaming. Jumping is a result of the reaction force from a rapidly growing droplet coalescence bridge impinging on the surface, breaking the oscillation symmetry.[42,43] The reaction from the surface from the normal impingement of the bridge has to be normal to it in the opposite direction. For roaming, there is no symmetry-breaking surface for a tangential reaction force to manifest as the condensate microdroplets are orders of magnitude larger than the underlaying nanostructures. If the increase in adhesion for all condensate droplets on the surface were uniform when subcooling increases, i.e. no asymmetry, jumping would gradually cease and transition to flooded condensation directly, without any intermediate in-plane roaming regime. The excess surface energy from coalescence would no longer overcome the increased adhesion and be dissipated instead. Droplets would not depart at all, whether in-plane or out-of-plane, until they attain the size when gravity dominates.

Moreover, any droplet size mismatch during coalescence cannot explain the generation of tangential momentum as well. In our roaming events, there is no observable trend in participating droplet sizes (**Figure 2a**). Additionally, for two size-mismatched coalescing droplets, the reaction force from the symmetry-breaking surface would still largely be normal to the surface. Numerical simulations (**Methods**) confirm that the direction of jumping from the coalescence of two size-mismatched droplets deviates < 4° from the surface normal, in line with previously reported results.[21] See **Supplementary Information S12**.

We therefore believe that roaming is a consequence of adhesion asymmetry, and propagates by the dewetting of the *some* partial-Wenzel state droplets as mentioned earlier. Current experimental methods do not simultaneously possess sufficient spatial and temporal resolution to visualise the pinned contact line of the adhered droplet, or the ≲100 nm-thick wetted nanostructure layer below the droplet, during coalescence. To study the generation of tangential momentum, we determine the evolution of momentum numerically (**Figure 4a** and **Methods**). Two equally sized droplets of 160 μm in diameter are first placed on a substrate at a contact angle of 160°, and a symmetry plane is specified at $z = 0$. To mimic the effect of a wetted nanostructure layer below a droplet (D1 in **Figure 4a**), we specify the contact angle only for its base area as 2°.

Contour plots of static gauge pressure at the symmetry plane are displayed in **Figure 4b**. In the beginning, the low pressure at the coalescence bridge draws the liquid to it which rapidly expands (Panel i). The span in the x-direction increases, followed by a recoil with a downward tendency due to the higher curvature at the +y end than the bottom. The recoil is asymmetric and biased towards -x as the higher wettability below Droplet D1 restricts liquid motion (Panel ii). This x-recoil in turn increases the span in the y- and z-directions. As the liquid body elongates in the y-direction, the adhesion of the hydrophilic wetted region below Droplet D1 creates a locally concave liquid-vapour interface (Panel iii). The pressure difference from the asymmetric curvature further draws the liquid towards -x. This cycle repeats, and the liquid body recoils in the other direction (z), biased towards -x, and experiences another curvature asymmetry (not shown in **Figure 4b**). See **Supplementary Information S13**. Maximum x-displacement is reached at ≈ 1 ms (Panel iv). In the current case, we do not specify any dewetting step. The liquid body then swings back towards +x.

The complete event for a duration of 2 ms is shown in **Video S6**, together with the reference case when both Droplets D1 and D2 are in the Cassie state. The evolution of momentum and centre of mass displacement in the x- and y-directions is plotted in **Figure 4c** against the reference case. In the latter, x-momentum and displacement remain zero whereas y-displacement increases continuously owing to droplet jumping. In the former, where the base of Droplet D1 is wetted, the absolute x-momentum reaches a maximum at ≈ 0.5 ms as the centre of mass of the system approaches the wetted region, ceasing tangential momentum generation. This mechanism reveals how wetting asymmetry due to increased adhesion of one droplet can serve as a hinge[21] to generate tangential momentum. Lastly, we vary the diameter and thus area of the wetted region as a percentage of the total base area below Droplet D1 and obtain the maximum tangential momentum generated for each percentage (**Figure 4d**). A sharp transition at ≈ 20% reveals that a slightly wetted base area can already generate substantial tangential momentum.

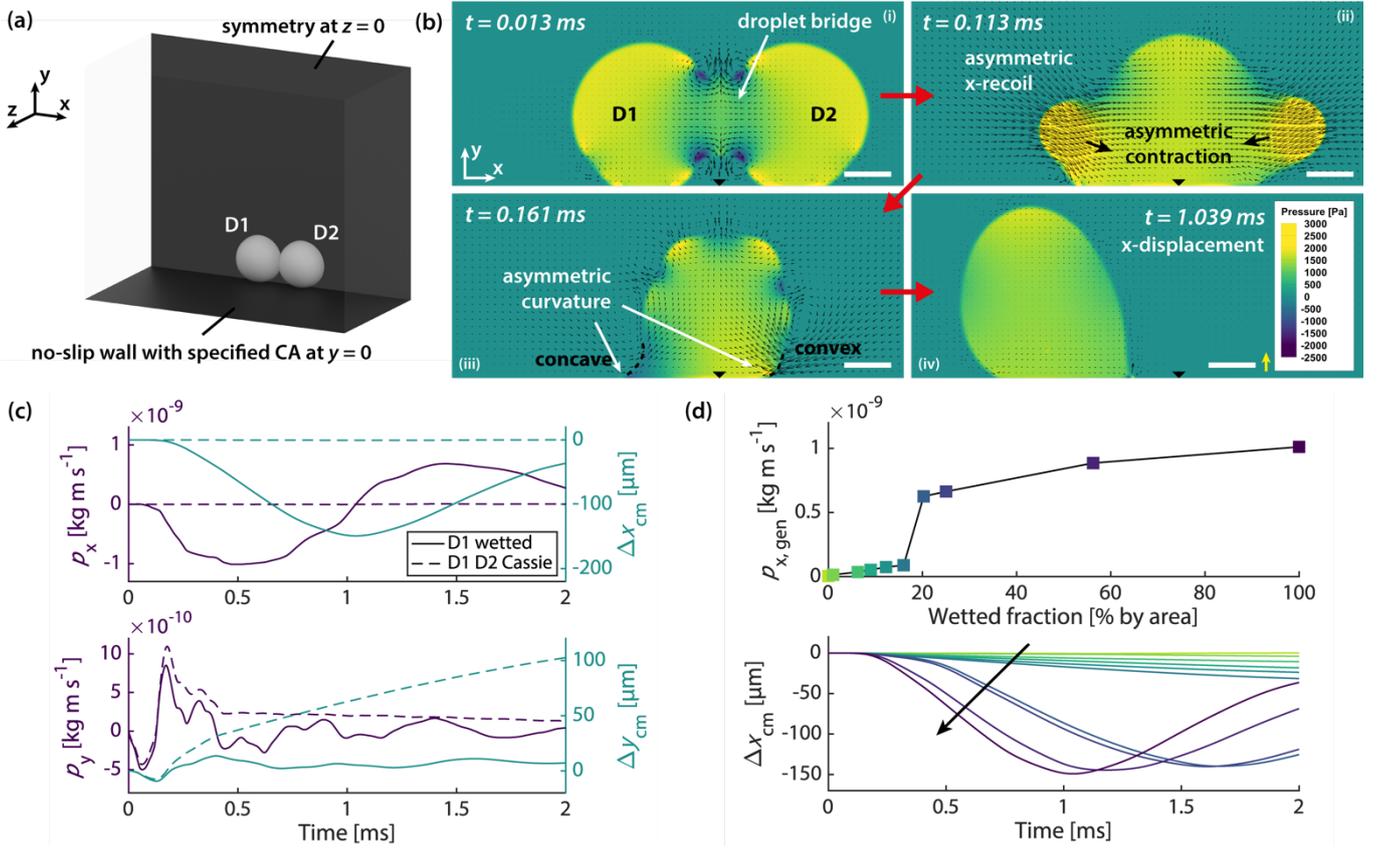

**Figure 4**: Generation of tangential momentum. (a) Computational domain. Two droplets with diameter 160 µm are placed on a no-slip wall at $y = 0$, specified with a contact angle. A symmetry plane is at $z = 0$. (b) Contour plots of static gauge pressure at the symmetry plane. The entire base area of Droplet D1 is wetted. Vectors are velocities. Scale bars: 50 µm. Yellow reference velocity vector: 2 m s$^{-1}$. (c) Momentum ($p_x$ and $p_y$ on the left y-axis) and centre of mass displacement ($\Delta x_{cm}$ and $\Delta y_{cm}$ on the right y-axis) in the x- and y-directions, for the case in which the base area of Droplet D1 is wetted and the case in which both Droplets D1 and D2 are in the Cassie state. (d) Maximum tangential momentum generated, $p_{x,\,gen} = \max(|p_x|)$, for varying wetted fractions of the base area of Droplet D1 (top), and the corresponding x-centre of mass displacement (bottom). In (c) and (d), the momentum reported reflects full spherical droplets, taking domain symmetry into account.

Dewetting and procession of roaming

After generating tangential momentum from asymmetric droplet adhesion, to be able to roam, the droplet has to dewet and detach from its location. In **Figure 5a**, we dewet the surface by reverting the specified contact angle of the wetted region (2˚) back to the original (160˚) at 179 μs (Panel i), when the force exerted on it in the +y-direction is at maximum. After recoiling in the z-direction (Panel ii), the coalesced droplet departs with a substantial tangential component (Panel iii). The departure angle is sensitive to the dewetting time, as dewetting 20 μs later already results in entirely tangential departure. In reality, the actual moment for dewetting depends on when both (1) static friction of the contact line[44] and (2) the adhesion work on the nanostructures are overcome. See **Video S7**.

We plot the variation of momentum and displacement in **Figure 5b** for the first 0.5 ms. After dewetting, tangential momentum generation ceases, but recovers in the normal direction. The kinetic energy of the translational motion of the centre of mass[43] compared to the total kinetic energy is expressed in **Figure 5c**. Although the total kinetic energy of our dewetting and the reference cases are similar, the former exhibits a higher centre-of-mass translational kinetic energy, indicating a higher efficiency in producing centre-of-mass motion instead of oscillatory viscous dissipation. See **Supplementary Information S14**.

Finally, we demonstrate the dewetting experimentally in **Figure 5d**. Coalescence can be seen at 0.2 ms. At 1.7 ms (Panel iii), the main droplet leaves slightly from the surface with a change in droplet reflection while maintaining significant tangential momentum. It is then intercepted by droplets on the surface and returns to the surface at 3.2 ms. In this event, we observe another dewetting at 5.5 ms (Panel v), before returning at 9.0 ms. See **Video S8** for more examples.

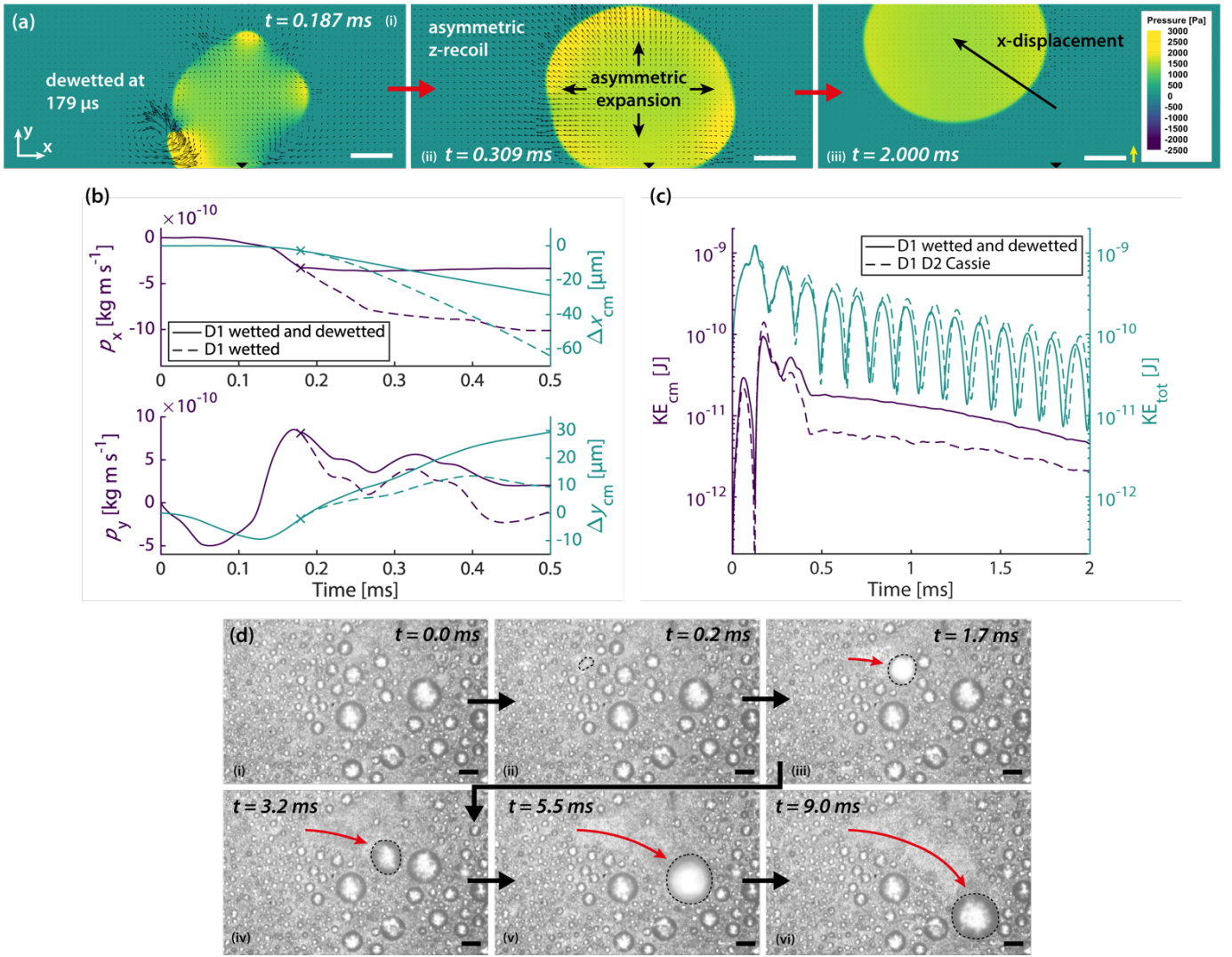

**Figure 5**: Dewetting in roaming. (a) Contour plots of static gauge pressure at the symmetry plane after dewetting at 179 μs. Initially the entire base area of Droplet D1 is wetted, similar to **Figure 4b**. Vectors are velocities. Scale bars: 50 μm. Yellow reference velocity vector: 2 m s$^{-1}$. (b) Momentum ($p_x$ and $p_y$ on the left y-axis) and centre of mass displacement ($\Delta x_{\text{cm}}$ and $\Delta y_{\text{cm}}$ on the right y-axis) in the x- and y-directions, for the case in which the original base area of Droplet D1 is subsequently dewetted at 179 μs, and the case in which it remains wetted. (c) Kinetic energy of the translational motion of the centre of mass $KE_{\text{cm}}$ and the total kinetic energy $KE_{\text{tot}}$ for the two cases. The momentum and kinetic energy reported in (b) and (c) reflect full spherical droplets, taking domain symmetry into account. (d) Experimental observation of dewetting as roaming progresses. Coalescence is seen at 0.2 ms (Panel ii). Dewetting is seen at 1.7 ms (Panel iii) and 5.5 ms (Panel v) as indicated by the change in reflection of the main droplet. Black dashed lines enclose the main droplet. Red arrow indicates the approximate trajectory of the roaming event. Subcooling: 2.0 K. Gravity is downwards. Scale bars: 100 μm.

**Conclusion**

The tangential momentum generation of coalescing condensate microdroplets on solely nanostructured superhydrophobic surfaces is attributed to the stochastic wetting state of the condensing droplets. The simultaneous presence of droplets at different wetting states results in adhesion asymmetry during coalescence, effectively converting excess surface energy to tangential kinetic energy as coalescence occurs. The ability of the coalesced droplet to dewet from the surface triggers roaming, while preventing condensate flooding of the surface. This frequently renews the surface for fresh nucleation. Remarkably, this process significantly improves heat transfer compared to other condensate removal modes, as it takes over as the dominant mechanism with jumping dropwise condensation subsiding at higher subcooling.

**Methods**

Formation of boehmite nanowalls[24–26]

All aluminium substrates are of EN AW-1050A. The substrates are sonicated in acetone, isopropanol, deionised water for 10 min respectively, followed by sonication in 0.25 M sodium hydroxide solution for at least 10 min, before rinsing with deionised water and drying with nitrogen. The samples are then placed in hot water at ≈ 96 ˚C for 10 min and dried with nitrogen.

pPFDA coating with iCVD[2,12]

The samples are first treated with oxygen plasma (Femto, Diener electronic) at 0.6 mbar for 10 min, followed by coating with trichlorovinylsilane (Sigma-Aldrich, CAS No.: 75-94-5) in a custom CVD chamber (saturated silane vapour at room temperature, ≈ 60 Torr). The samples are then placed in an iCVD system (iLab, GVD), to form a pPFDA coating at 100 mTorr using tert-butyl peroxide (Sigma-Aldrich, CAS No.: 110-05-4) as the initiator and 1H,1H,2H,2H-perfluorodecyl acrylate (Sigma-Aldrich, CAS No.: 27905-45-9) as the monomer. The stage and filament temperatures are set to 40 ˚C and 300 ˚C respectively. When applied on a pristine silicon wafer, the coating gives an advancing contact angle, contact angle hysteresis, and static contact angle of 124.2˚ ± 0.4˚, 12.5˚ ± 2.1˚ and 119.2˚ ± 1.5˚ respectively.

Contact angle goniometry

Advancing and receding contact angles are measured with a goniometer (OCA 35, DataPhysics Instruments). Deionised water is deposited and withdrawn at a rate of 2 µL s$^{-1}$. The sample is blown dry with nitrogen before deposition of every droplet. Three measurements are taken before and after condensation. No significant change in wettability is observed. The static contact angle is computed from the mean of advancing and receding contact angles as a single droplet cannot be stably deposited.

Numerical simulations

Cases are set up and computed with Ansys ICEM CFD and Ansys Fluent using the volume of fluid method. Saturation properties at 30 mbar are specified for the fluids. Postprocessing is performed in Tecplot 360 EX and MATLAB (MathWorks). See **Supplementary Information S15** for details.

**Acknowledgements**

We thank Tobias Neef for his assistance with the iCVD process, and Jovo Vidic and Peter Feusi for their assistance in the construction of the condensation setup. We thank Henry Lambley and Jonathan Boreyko for helpful discussions; and Thibaut Delafosse and Mithulan Vasan for assisting in preliminary experiments. We thank Jiayu Song for preparing the titanium samples. We thank the Cleanroom Operations Team of the Binnig and Rohrer Nanotechnology Center (BRNC) for their help and support. Unless otherwise specified, fluid properties are obtained with CoolProp (www.coolprop.org).[45] This project has received funding from the European Union's Horizon 2020 research and innovation programme under grant number 801229 (HARMoNIC).

**Author contributions**

C.W.E.L. and D.P. conceived the research. D.P. supervised all aspects of the research and provided scientific guidance. C.W.E.L. designed and constructed the condensation setup, conducted the experiments, performed the simulations, and analysed the data. C.W.E.L. prepared the aluminium samples. M.D. prepared the copper samples. C.W.E.L., K.R., and A.T. applied the pPFDA coatings. G.C.P. assisted in the simulations. D.P., K.R., C.S.S., and A.M. provided scientific guidance for the various aspects of the research. C.W.E.L. and D.P. wrote the manuscript with contribution from all other authors.

**Supplementary information for**

# Condensate droplet roaming on nanostructured superhydrophobic surfaces

Cheuk Wing Edmond Lam[1], Kartik Regulagadda[1], Matteo Donati[1], Abinash Tripathy[1], Gopal Chandra Pal[2], Chander Shekhar Sharma[2], Athanasios Milionis[1], and Dimos Poulikakos[1, *]

[1] Laboratory of Thermodynamics in Emerging Technologies, Department of Mechanical and Process Engineering, ETH Zurich, Sonneggstrasse 3, 8092 Zurich, Switzerland

[2] Thermofluidics Research Lab, Department of Mechanical Engineering, Indian Institute of Technology Ropar, Rupnagar, Punjab, 140001 India

* Corresponding author
Prof. Dr. Dimos Poulikakos
Email: dpoulikakos@ethz.ch
Phone: +41 44 632 27 38
Fax: +41 44 632 11 76

**Description of supplementary videos**

Video S1: Typical roaming event

Video S2: Local clustered multi-droplet coalescence

Video S3: Roaming events altering direction

Video S4: Roaming events terminating in jumping

Video S5: Condensation modes and heat transfer coefficients at different subcooling

Video S6: Source of tangential momentum

Video S7: Motion for dewetting at different times

Video S8: Dewetting during roaming

**Table of contents**

## S1. Thickness of the pPFDA layer by ellipsometry

The thickness of the pPFDA layer is measured by ellipsometry (V-VASE, J.A. Woollam), using a reference silicon wafer coated with the same iCVD process as the samples. Specifying the material stack, data fitting is performed with the software provided by the company to determine the layer thickness. Measurement is taken for at least two locations on each sample for the mean, over wavelengths of 200 – 1600 nm for at least three different angles.

The native oxide thickness of silicon is first determined to be 1.7 nm with a material stack of silicon (525 µm) and silicon dioxide, using a sample cut from a pristine silicon wafer. Next, the pPFDA layer is added to the material stack as a Cauchy material, using constants A: 1.3992, B: 0.0069215, C: -0.00024462, which are found to provide the best fit. The thickness of the pPFDA layer is determined to be 3.5 nm. An example for the data fitting at one location can be found in **Figure S1.1**.

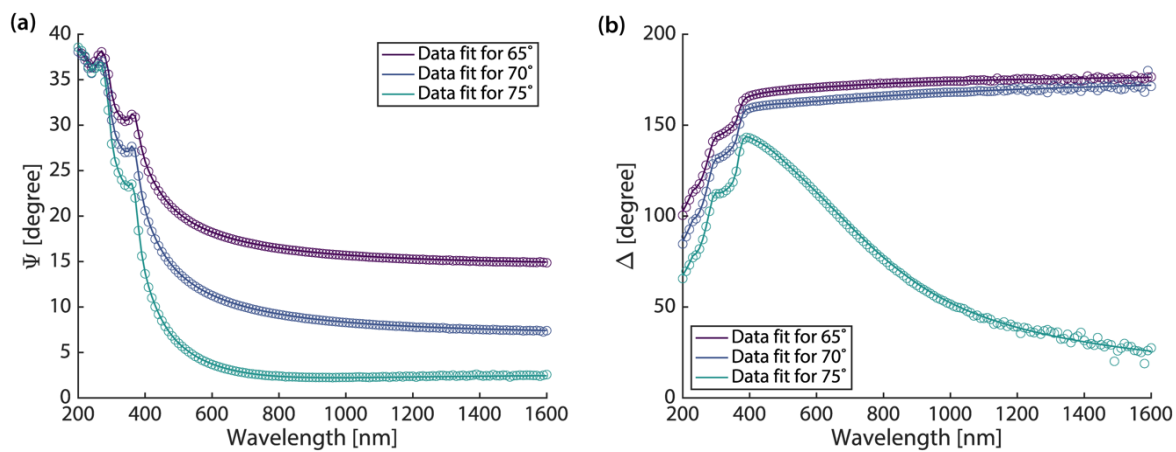

**Figure S1.1**: Raw ellipsometer measurements (circles) and their fit to the model (solid lines) at one location of a pPFDA-coated silicon wafer. Each colour represents measurement at one angle. (a) Spectra of the amplitude component Ψ and (b) the phase difference Δ.

## S2. Condensation and observation setup

Condensation setup

The condensation setup is similar to our previous work.[1,2] Condensation is performed with saturated steam at a nominal pressure of 30 mbar. The condensation chamber is installed as a part of an open system consisting of a pressure and steam source, the boiler, and a pressure sink, the vacuum pump. The schematic of the system is shown in **Figure S2.1**.

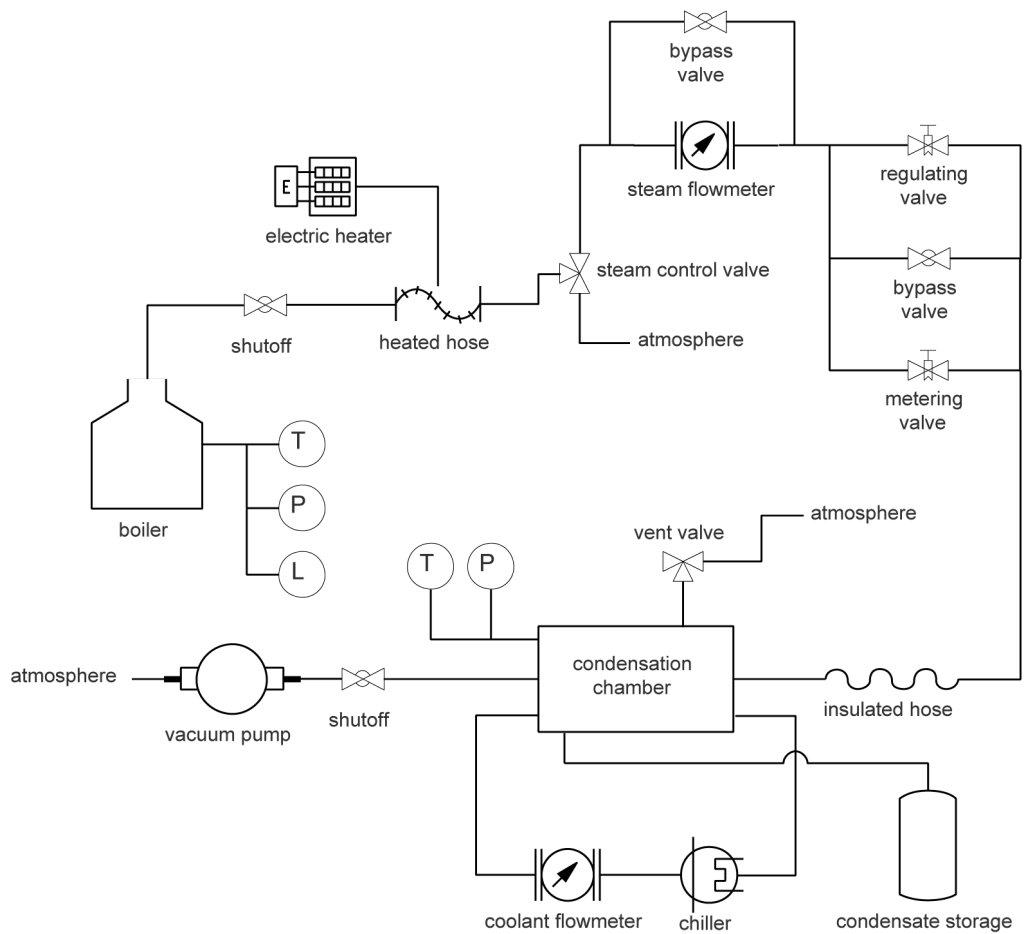

**Figure S2.1**: System consisting of the condensation chamber. Reproduced from our previous work[2] under Creative Commons licence CC-BY-NC-ND.

During operation, a boiler containing deionised water is used to generate steam at 1.4 ± 0.01 bar. The steam passes through a heated hose, and its flow rate is measured by a flowmeter (FAM3255, ABB). Regulating (SS-6BMRG-MM, Swagelok) and metering (SS-6BMW-MM, Swagelok) valves are used to control the steam flow so that the steam in the condensation chamber is at saturation with a pressure of 30 mbar. The steam passes through an insulated hose and enters the chamber, in which the sample is mounted and cooled by a recirculating chiller. Flow rate of the chiller is monitored by a flowmeter (SITRANS FM MAG5000 and SITRANS FM MAG 1100, SIEMENS). In the chamber, the steam pressure and temperature, as well as the sample surface temperature, are continuously measured at 2 Hz. All other sensors of the system are connected to the same data acquisition device and measured at 2 Hz (Beckhoff). Excess condensate is collected at the storage. At the exit of the chamber, a vacuum pump (RC 6, VACUUBRAND) is used to drive a stable steam flow.

A schematic of the chamber is found in **Figure S2.2**.

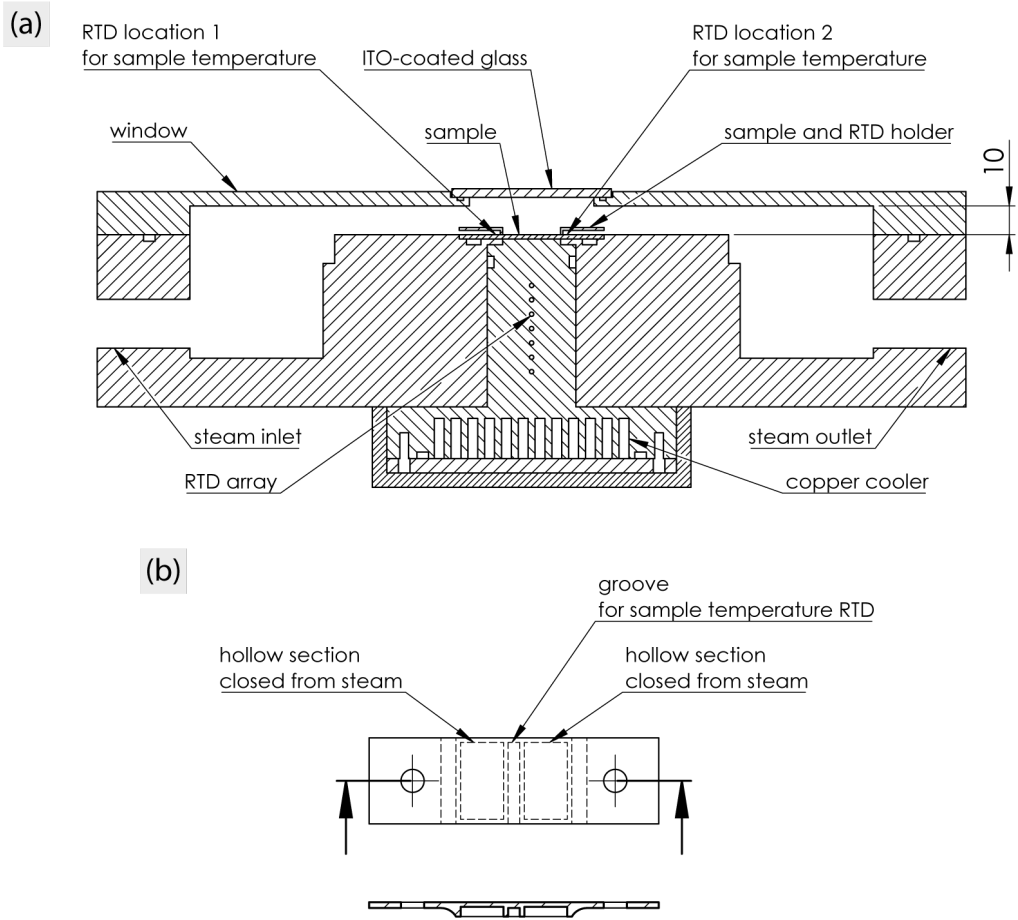

**Figure S2.2**: Top-view cross section of the condensation chamber. Dimension in mm. Reproduced from our previous work[2] under Creative Commons licence CC-BY-NC-ND.

The condensation chamber is where the sample is mounted and tested. The base of the chamber is milled from a block of polyether ether ketone (PEEK), and its front window is milled from a block of poly(methyl methacrylate) (PMMA).

Refer to **Figure S2.2a**. The sample of size 50 mm × 20 mm is mounted vertically in the middle of the chamber, on one end of the copper cooler, which measures 20 mm × 20 mm in size. During operation, only the centre 20 mm × 20 mm region of the sample directly on top of the cooler is exposed to steam, and the remaining two 15 mm × 20 mm side regions of the sample are insulated from direct steam exposure. The insulation is achieved by the design of the mounts (**Figure S2.2b**). On top of each side region, a 3D-printed hollow polycarbonate mount is screwed into the chamber to provide pressure onto the sample and fix it onto the cooler. Below each mount, a groove is present to accommodate a Pt 1000 Class A resistance temperature detector (RTD) (P1K0.516.1K.A.152.D.S, IST), which is attached to the sample with Kapton tape for measuring the surface temperature. The mean of the two surface temperature RTDs is taken as the measured value.

For steam conditions in the chamber, the steam pressure is measured with a capacitance gauge (CMR 362, Pfeiffer Vacuum) and the steam temperature is measured with two Pt 1000 Class A RTDs (P1K0.516.1K.A.152.D.S, IST). Similarly, the mean of the two steam temperature RTDs is taken as the measured value.

The sample is fixed by the mounts onto the front end of the cooler. Between the sample and the cooler, a thermal paste (KP 99, Kerafol) is applied. The cooler is milled from a block of copper (CW004A, Durofer). The back end of the cooler is a heat exchanger with a coolant recirculated by a chiller (WKL 2200, LAUDA). Between the front and back ends of the cooler, an array of 7 Pt 100 Class A RTDs (Thermo Sensor) is used to determine the temperature gradient and the heat flux using the thermal conductivity of the cooler (394 W m$^{-1}$ K$^{-1}$). For this part of the cooler where the 7 RTDs are located, there is a thin closed air gap between the cooler and the PEEK chamber base as extra insulation (not drawn in **Figure S2.2a**).

The temperature of the back end of the cooler is directly controlled by the temperature of the coolant, which is input to the chiller. Given a temperature difference between the steam (fixed at saturation temperature, 24.1 ˚C) and the back end of the cooler (varied between 20 and -10 ˚C), the tested condensing surface determines its resulting subcooling and the heat flux through it. Better-performing surfaces result in a smaller range of resulted subcooling, as less thermal resistance is present between the surface and the steam.

At the centre of the PMMA window, an indium-tin-oxide-coated borosilicate glass (Diamond Coatings) is installed into a cut-out. A slight voltage is applied to the coating to provide minimal heating to remove condensate fogging as necessary.

To estimate the amount of non-condensable gases in the chamber during operation, we measure the inward leakage rate. The chamber is first evacuated by the vacuum pump overnight to achieve the minimum attainable pressure (< 1 Pa). The connection to the pump is then turned off and the chamber is left to leak inwards. We perform three leakage tests and obtain an estimated leakage rate of ≈ 0.09 – 0.22 mbar h$^{-1}$. We therefore determine that the amount of non-condensable gases is negligible in our experiments. We further note that in our experiments, there is a constant supply of fresh steam from the boiler and the vacuum pump is constantly running as well. Steam continuously flows across the test surface (**Figure S2.2a**), and non-condensable gases do not accumulate in the chamber.

Observation setup

The observation setup is similar to our previous work.[3] Refer to **Figure 1b** in the main text. A microscope objective (UPlanFl 4x/0.13 PhL, Olympus) is placed in front of the chamber window. White light is generated by an LED (LEDD1B and MCWHF2, Thorlabs) and carried through an optical fibre (QP1000-2-UV-BX, Ocean Optics) to the optical assembly. A set of lens adjusts the incoming beam size before it reaches the beam splitter. The beam splitter reflects the incoming light to the microscope objective, after which a circular region of ≈ 4 mm in diameter on the sample is illuminated. The reflected light from the sample passes through the microscope objective and the beam splitter to another set of lens to focus onto the image sensor of the high-speed camera (FASTCAM SA1.1, Photron). With a reflective surface such as boehmite, a frame rate of 10,000 fps can be achieved.

Experimental procedures

We follow similar experimental procedures to our previous work.[1,2] For each experiment, the chamber is first pumped at least overnight by the vacuum pump with an ultimate pressure of 0.002 mbar. The bypass valve parallel to the steam flowmeter and the bypass valve parallel to the regulating and metering valves are open, and the 3-way steam control valve before the steam flowmeter is closed. The components between the steam control valve and the vacuum pump are thus pumped. Pumping overnight allows the removal of all condensate from previous experiments remaining in these components. The chamber then reaches its minimum pressure, which is below our measurement limit of 0.01 mbar.

After one night, the chamber is vented, and the tested sample is mounted onto the cooler with thermal paste as described above. The chamber is pumped down again, and the pump runs continuously from this point to the end of the experiment.

The chiller is set to a temperature of 25 °C, slightly higher than the target saturation temperature of 24.1 °C, such that when steam is later introduced into the chamber, condensation does not immediately occur. Flow rate of the coolant is set to $180 \pm 10$ L h$^{-1}$. The boiler is filled with deionised water, set to be open to atmosphere using the 3-way steam control valve, and turned on. The water is boiled at > 1.4 bar for 30 min to degas. At the same time, the two bypass valves are cycled open-and-close multiple times to release the trapped gases in them.

After 30 min of degassing, the two bypass valves are closed and the steam control valve is switched from the atmospheric side to the chamber side. Steam is introduced into the chamber. The LED light source is turned on and set to maximum until the end of the experiment. Using the regulating and metering valves, the steam pressure in the chamber is slowly increased to the target pressure of 30 mbar. At the same time, boiler power is adjusted to maintain a pressure of 1.4 ± 0.01 bar until the end of the experiment. As the chamber pressure approaches 30 mbar, the chiller is set to the first set point of 20 ˚C. Condensation takes place as the surface temperature reaches below the chamber steam temperature. At this stage, the steam in the chamber has already reached saturation. Fluctuations in the measured chamber steam temperature can be seen to closely follow the fluctuations in the measured chamber steam pressure.

When the chiller reaches 20 ˚C, the flow rate is readjusted to 180 ± 10 L h$^{-1}$ to compensate for the change caused by thermal contraction of the coolant. The system is then adjusted to stabilise by fine-tuning the boiler power and the metering valve, until it can maintain steady-state conditions without intervention.

The glass window is then checked for condensate fogging. If present, a voltage of 10 V is applied to the indium-tin-oxide coating at a current of ≈ 580 mA for 15 s to generate minimal heat to remove the condensate fog on the window. No measurable increase in chamber steam temperature by this heating is detected. The microscope objective is refocused. The system is then left without intervention for 1 min, and the measurements within this minute are used to compute the steady-state data point (subcooling, heat flux and heat transfer coefficient) at this set point. During this minute, the chamber pressure has to maintain at 30 ± 0.5 mbar and the boiler pressure has to maintain at 1.4 ± 0.01 bar without intervention. Approximately 30 s into this minute, a video is taken with the high-speed camera at 10000 fps for ≈ 1 s, which is the condensation behaviour attributed to this data point. The procedure in this paragraph is repeated when more than one video is desired for this chiller set point.

The chiller is then set to proceed to the next set point. For the results presented in **Figure 3**, a total of 7 set points are set, from 20 to -10 ˚C at 5 ˚C intervals. As the chiller reaches the next set point, the procedures above are repeated, i.e. coolant flow rate readjustment, system stabilisation, condensate fogging check, refocusing, steady-state measurement and high-speed video recording.

After the data point at the last chiller set point (-10 ˚C) is complete, the chiller is set to 25 ˚C. The boiler is turned off, but steam continuously passes over the sample as there is still a pressure difference between the boiler and the pump, preventing the surface from reaching freezing temperatures as the chiller takes time to heat up.

The LED light source is turned off. When the coolant temperature reaches 5 ˚C, the chiller is turned off. The 3-way steam control valve from the boiler, and the shutoff valve to the pump, are closed, stopping the steam flow. The chamber is vented, and the sample is removed from the cooler and blown dry with a weak nitrogen flow.

Computation of heat fluxes and heat transfer coefficients

Heat flux is computed from the linear temperature gradient measured by the RTD array in the cooler **(Figure S2.2a)**. The one-dimensional Fourier's law of conduction states:

$$q = kA\frac{dT}{dx}$$

where $k = 394$ W m$^{-1}$ K$^{-1}$ is the thermal conductivity of the cooler, $A$ is the cross-sectional area of the cooler, and $dT/dx$ is the temperature gradient obtained from the linear fit of the 7 measured temperatures by the array. $q$ is the heat flow through the cooler. As the cooler is insulated with a closed air gap and a thermally insulative PEEK chamber base, the heat flow through the cooler can be assumed to be equal to the heat flow through the condensing surface, i.e. there are negligible thermal losses. Therefore, the heat flux of the condensing surface $q''$ can be computed by dividing the heat flow $q$ by its area, $q'' = q/A$, where $A = 20$ mm × 20 mm = $4 \times 10^{-4}$ m$^2$ is the area of the condensing surface instead.

The heat transfer coefficient $h$ is defined as:

$$h = \frac{q''}{\Delta T}$$

where $\Delta T = T_{\text{steam}} - T_{\text{surf}}$ is the subcooling, i.e. the temperature difference between the measured steam temperature $T_{\text{steam}}$ and the measured surface temperature $T_{\text{surf}}$.

Every data point represents steady-state measurements of a 1-min period. In **Figure 3a** and **S8.1**, for each surface, 7 data points are plotted, reporting the heat transfer coefficients and heat fluxes for 7 subcooling. The temporal mean of the measured subcooling in this 1-min period (2 Hz, 120 instantaneous values) is the reported subcooling for this data point. Similarly, the reported heat flux and heat transfer coefficient for this data point is the temporal mean of the respective values in this 1-min period.

Error bars reported in **Figure 3a** and **S8.1** are calculated from both the uncertainty of the sensors as well as the fluctuations of the measurements. Refer to our previous work[2] for the error propagation procedure.

**S3. Effect of microstructures on droplet motion**

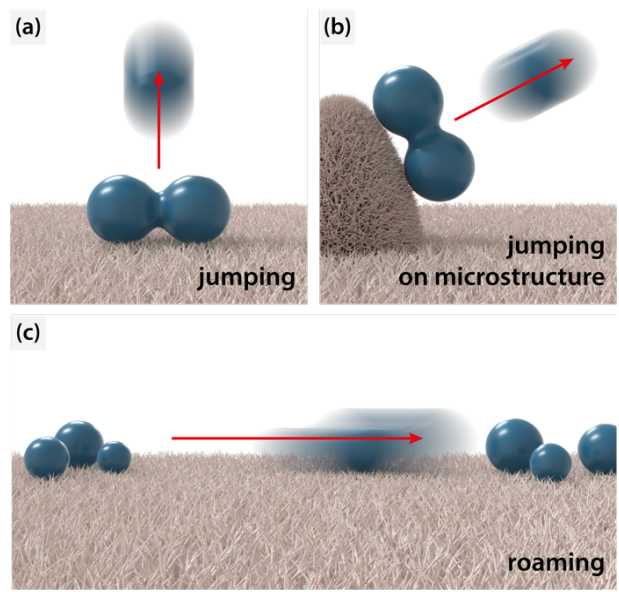

**Figure S3.1**: Different droplet motion behaviour. (a) Motion from droplet jumping is normal to the surface. (b) When there are microstructures which are at the same length scale as the droplets, the direction of jumping can be altered. (c) Roaming can occur without microstructures.

**S4. Image processing, droplet measurement and event tracking**

High-speed videos in the form of image sequences are first processed in Adobe Lightroom Classic to enhance contrast. The resulting image sequence is imported into ImageJ (National Institutes of Health) for further processing. The measurement of participating droplets and the main droplet are used to study the evolution and mechanics of roaming events.

Participating droplets

Participating droplets refer to all the droplets which coalesce in a roaming event. Before coalescence, they possess a spherical cap geometry. Using the oval tool in ImageJ, the centre and diameter of the circular participating droplets are measured. They are measured at the frame (time) before the beginning of their coalescence, determined by a visible perturbation to their interface. Therefore, a time is assigned to every participating droplet. The time of the first participating droplet is set to $t = 0$ and the time of every other is set relative to this first droplet. The duration of the roaming event is the duration between the first and the last participating droplet coalescence.

For participating droplets much smaller than the main droplet, their coalescence with the main droplet cannot be seen as they may lie under the main droplet at the point of coalescence. However, their coalescence with the main droplet can be confirmed by the absence of these participating droplets after the main droplet traverses and leaves behind a clear renewed area. In these cases, the participating droplets are measured at the frame when they are last seen.

The directionality of roaming events can be seen in the spatial distribution of participating droplets. If a convex hull is drawn around the participating droplets, we observe a lower circularity ($= 4\pi(\text{area}/\text{perimeter}^2)$) of the hull for roaming events ($0.70 \pm 0.08$) than the hull for clustered multi-droplet coalescence ($0.89 \pm 0.06$).

Main droplet

The main droplet refers to the traversing droplet, resulted from the coalescence of previous participating droplets, which sweeps and coalesces with the remaining participating droplets until termination of the roaming event. The main droplet has an irregular geometry. The contour of their projected shape on the condensing surface is measured using the polygon tool in ImageJ. The centroid of the measured polygon is the location of the main droplet at that time.

We begin to measure the main droplet when its contour can be clearly seen for the first time, and measure the main droplet for each frame afterwards. For most roaming events, the main droplet comes to rest on the surface at the end of the event. In this case, the main droplet is measured until 20 frames (2 ms) after the time of the last participating droplet. Occasionally, the main droplet jumps and departs from the surface to terminate the roaming event. In this case, the main droplet is measured until 1 frame (0.1 ms) after the time of the last participating droplet.

Main droplet contours measured in ImageJ are imported into MATLAB for processing using code on the MATLAB Central File Exchange.[4]

Additional notes on measurement

The measurement of participating droplets and the main droplet are used to study the evolution and mechanics of roaming events. Therefore, longer events which proceed in all in-plane directions are chosen for measurement. We also avoid measuring roaming events which are blocked by a substantially larger droplet (compared to the size of the main droplet before coalescence with it), which brings the event into an abrupt stop as the tangential momentum is largely dissipated after the main droplet coalesces into it. Every measured event begins spontaneously, i.e. there is no incoming landing droplet to transfer momentum, and all participating droplets are at rest before the beginning of the event without remnant oscillations from previous droplet motion events.

It is noted that there are inherent difficulties in the extraction of the behaviour of the main droplet. Discontinuities arise as it traverses and coalesces with participating droplets. For example, at the moment of coalescence with a participating droplet, the main droplet instantaneously gains in projected area, resulting in a sudden movement in the location of its centroid and the velocity of the main droplet cannot be well defined. Similarly, jumps occur in the main droplet mass as participating droplets coalesce into the main droplet. Discontinuities in the velocity and mass prohibit reliable computation of the momentum and kinetic energy of the main droplet.

**S5. Roaming on various nanostructured surfaces**

Roaming has been reported on silicon nanograss (black silicon)[5] and copper nanowires[6] in the literature. We demonstrate its generality on different nanostructures with observation on titanium dioxide nanorods[7] and copper(II) hydroxide nanoneedles[8] in this section. It is noted that due to the poor reflectivity of these samples, the image quality is significantly lower than boehmite in **Figure 3b**. Nevertheless, condensation modes can still be identified.

Titanium dioxide nanorods

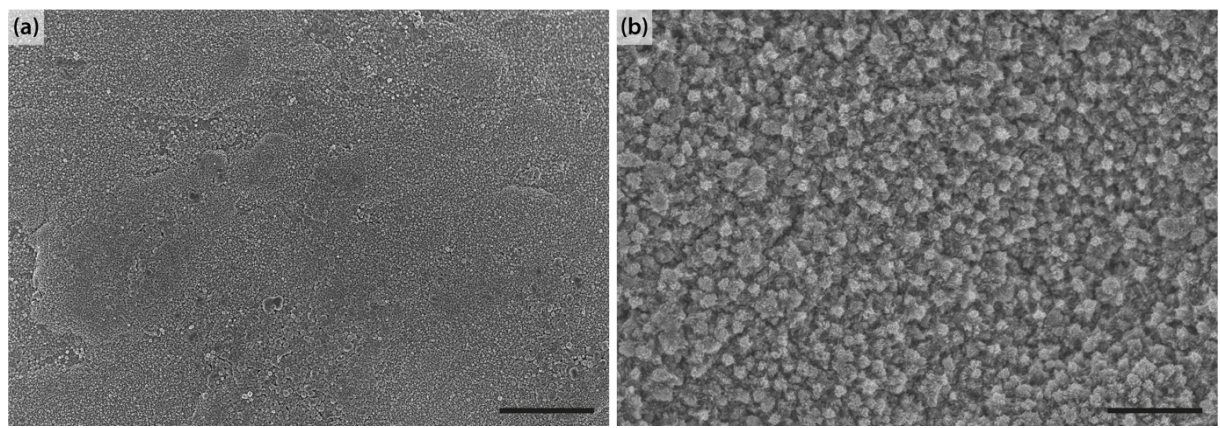

**Figure S5.1**: SEM images of titanium dioxide nanorods with the silane coating at different magnifications. (a) No prominent structures are present at the microscale. Scale bar: 20 μm. (b) Structures at the nanoscale. Scale bar: 2 μm.

Branched nanorod structures are prepared on flat titanium substrates. These structures are nano-hierarchical, i.e. both the larger-scale and smaller-scale structures are at the nanoscale. No structures are in the microscale to affect jumping direction as in **Figure S3.1b** found on conventional micro-nano hierarchical surfaces. The structured surface is then functionalised with 1H,1H,2H,2H-perfluorodecyltrichlorosilane. Fabrication procedures can be found in our recent work.[7] In the following, we expose samples with branch lengths of ≈ 90 nm (type b-NR-90 as specified in the work[7]) to the same condensation conditions as boehmite. SEM images of the structures can be found in **Figure S5.1**.

These surfaces are superhydrophobic with a measured advancing contact angle of 159.5° ± 1.3° and contact angle hysteresis of 1.8°. Roaming is seen when subcooling is increased. See **Figure S5.2**.

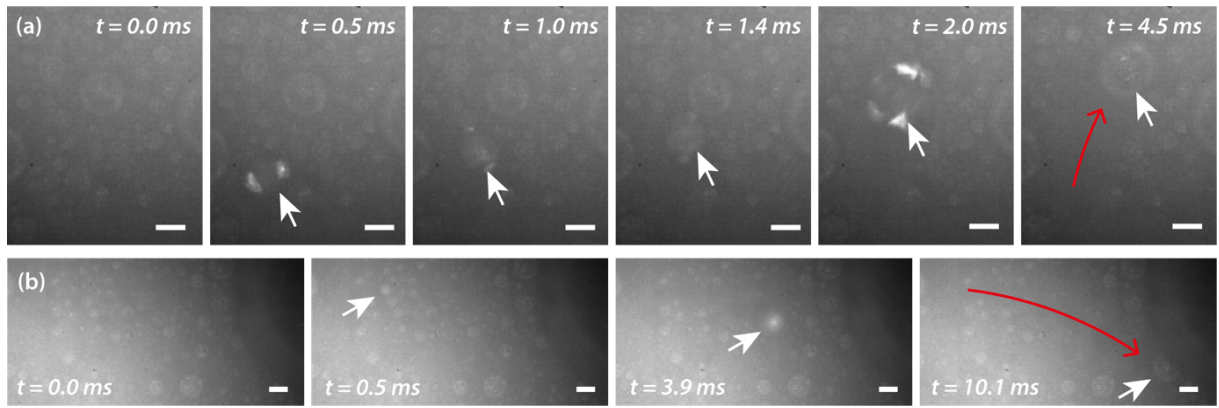

**Figure S5.2**: Roaming on titanium dioxide nanorods. White arrow points to the main droplet. Red arrow indicates approximate trajectory. (a) Upward event against gravity. Subcooling is at 2.7 K. Scale bars: 100 μm. (b) Rightward event. Subcooling is at 1.7 K. Scale bars: 100 μm.

Copper(II) hydroxide nanoneedles

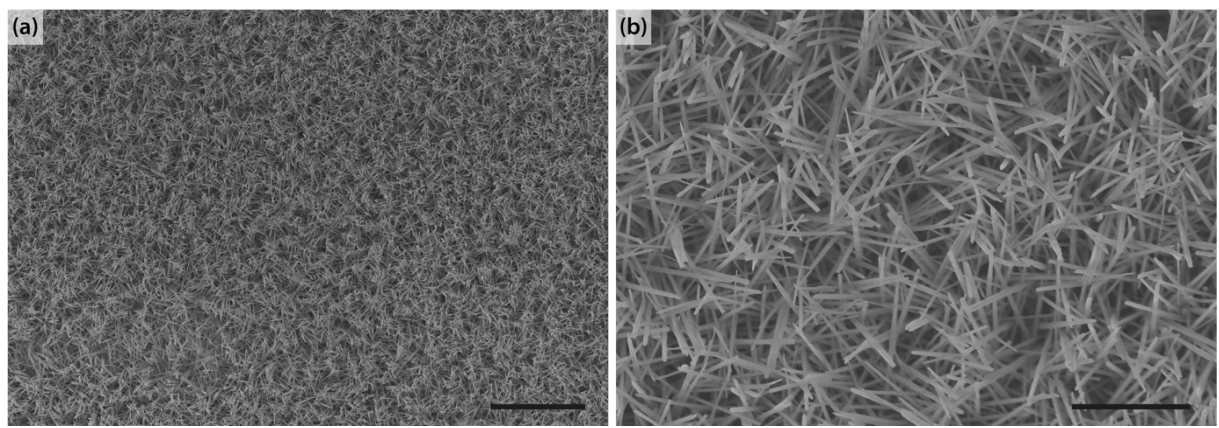

**Figure S5.3**: SEM images of copper(II) hydroxide nanoneedles with the pPFDA coating at different magnifications. (a) No prominent structures are present at the microscale. Scale bar: 20 µm. (b) Structures at the nanoscale. Scale bar: 5 µm.

Nanoneedle structures are prepared on flat copper substrates according to procedures from our recent work,[8] and then functionalised with pPFDA using iCVD (**Methods**). Similarly, there are no structures at the microscale. SEM images of the structures can be found in **Figure S5.3**. We expose these surfaces to the same condensation conditions as boehmite.

These surfaces are superhydrophobic with a measured advancing contact angle of 160.2° ± 1.3° and contact angle hysteresis of ≈ 0°. Roaming is seen when subcooling is increased. See **Figure S5.4**.

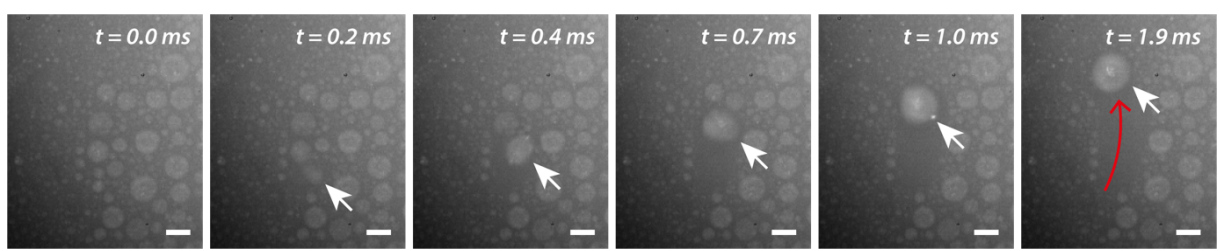

**Figure S5.4**: Upward roaming on copper(II) hydroxide nanoneedles against gravity. White arrow points to the main droplet. Red arrow indicates approximate trajectory. Subcooling is at 0.7 K. Scale bars: 100 µm.

## S6. Evolution of roaming events

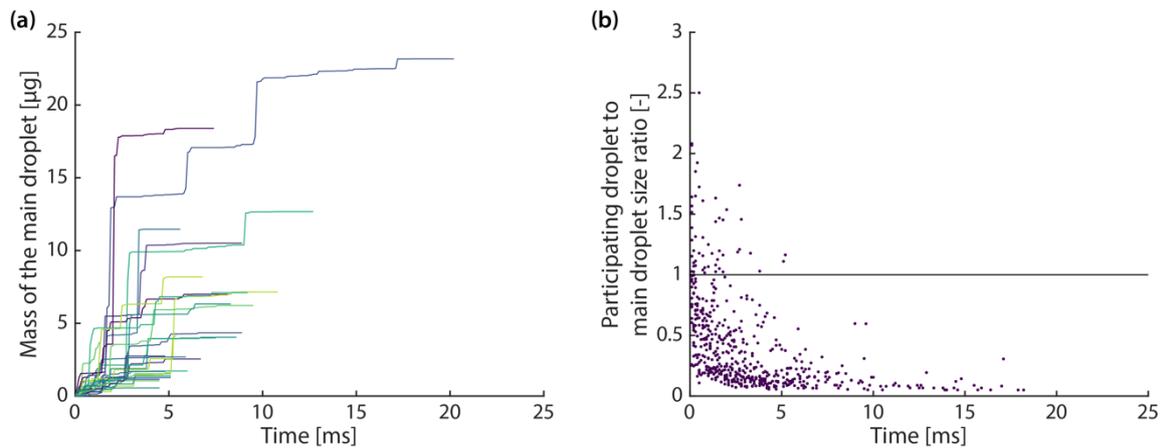

**Figure S6.1**: (a) Mass of the main droplet over time. (b) Ratio of participating droplet sizes relative to the main droplet size, from all measured roaming events.

Assuming all droplets exhibit a spherical cap shape and a constant contact angle (= 162.0°) with the surface, the volume of each participating droplet can be estimated from the measured diameter. The volume of the main droplet at any given time can then be estimated by summing the volume of all participating droplets which have coalesced at this time. The mass of the main droplet is computed by multiplying the volume with the density. **Figure S6.1a** shows the increase in the mass of the main droplet over time for different roaming events. The mass gain of each event varies significantly.

In a roaming event, as the main droplet gains in size, the size of participating droplets relative to it reduces. In **Figure S6.1b** we compute the ratio of participating droplet sizes relative to the main droplet size. The ratio is computed as follows: 1) The volume of the main droplet at each time point is converted to an equivalent diameter as if it assumed a spherical cap shape and a contact angle of 162.0°; 2) For each time point, if there is at least one participating droplet coalescing, the ratio at this time point is computed by dividing the diameter of the largest participating droplet coalescing, by the equivalent diameter of the main droplet. It can be seen in **Figure S6.1b** that this ratio drops below unity at ≈ 5 ms.

## S7. Real roaming velocity

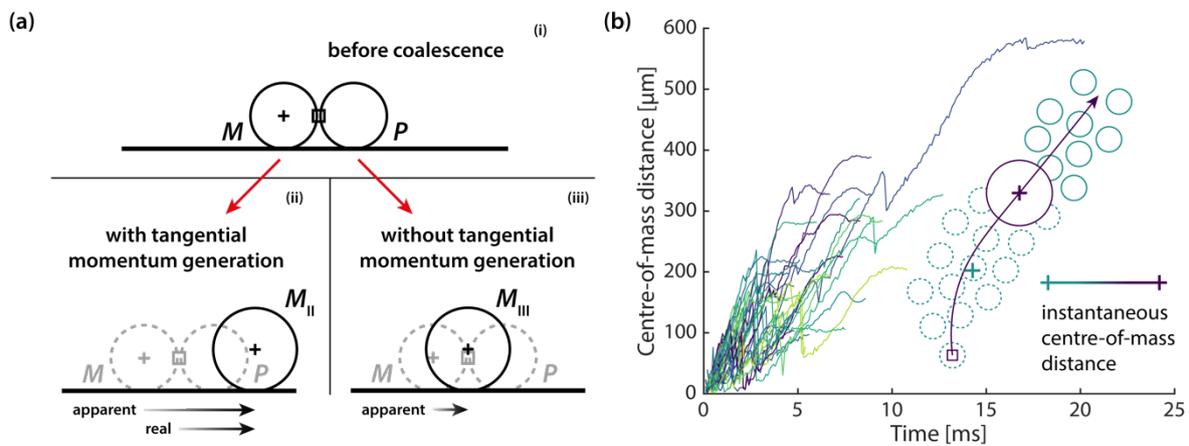

**Figure S7.1**: (a) Shift in the centre of mass of the main droplet upon coalescence, with and without tangential momentum generation. Cross: Centre of mass of the main droplet. Square: Centre of mass of the system of droplets before the coalescence (*M* and *P*). (b) Evolution of the distance between the centre of mass of the main droplet and the centre of mass of the system of participating droplets for the measured roaming events. Schematic illustrates a roaming event in progress. Square: Starting location of the roaming event. Violet circle: main droplet. Turquoise circles: coalesced (dashed) and to-be-coalesced (solid) participating droplets. Violet cross: Centre of mass of the main droplet. Turquoise cross: Centre of mass of the system of coalesced participating droplets (dashed turquoise circles).

**Figure S7.1a** illustrates the shift in the centre of mass of the main droplet upon coalescence. The original location of the main droplet *M* and a participating droplet *P* is shown in Panel (i). Upon coalescence, two outcomes can occur for the location of the new main droplet, $M_{II}$ in Panel (ii) or $M_{III}$ in Panel (iii). If there is tangential momentum generation (Panel (ii)), the location of the new main droplet $M_{II}$ will be away from the centre of mass of the system of droplets before the coalescence (*M* and *P*), indicated by the square. In contrast, if there is no tangential momentum generation (Panel (iii)), the location of the new main droplet $M_{III}$ will be unshifted from the centre of mass of the system of *M* and *P*.

However, even if there is no tangential momentum generation (Panel (iii)), it is clear that there is a displacement of the location of the main droplet from $M$ to $M_{III}$, due to the addition of the mass of $P$ to $M$, away from the location of $M$. This displacement is apparent, and not a result of tangential momentum generation. Tracking the location of the main droplet results in the apparent roaming velocity as described in the main text.

To isolate the result of tangential momentum generation, we compute the distance between the current centre of mass of the main droplet at every time instant of a roaming event and the centre of mass of the system of participating droplets which have coalesced by that instant. See the schematic in **Figure S7.1b**. A roaming event in progress is illustrated. The distance between the centre of mass of the main droplet (violet cross) and the centre of mass of the system of participating droplets which have coalesced (turquoise cross) is measured over time of the roaming event. The measurements of several roaming events are individually plotted in **Figure S7.1b** with different colours. The main droplet is consistently "ahead" of the system of coalesced participating droplets at any given time. An increase in the distance over time indicates a continuous tangential momentum generation in one in-plane direction.

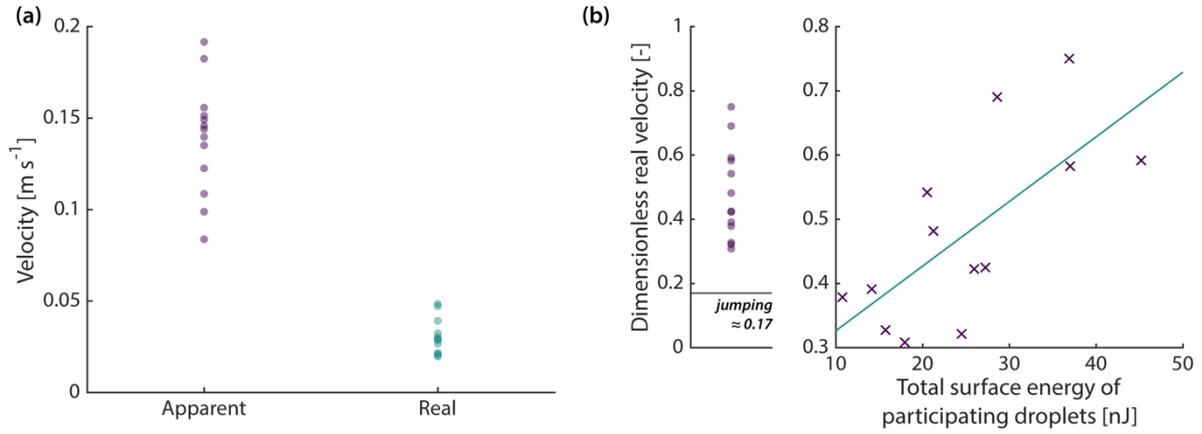

**Figure S7.2**: (a) Apparent roaming velocity and real roaming velocity at 5 ms. (b) Left: distribution of the dimensionless real velocity of roaming. For jumping, the dimensionless velocity is 0.17.[9] Right: dimensionless real velocity against the total surface energy of participating droplets.

To extract the characteristic real roaming velocity as a result of tangential momentum generation, we consider the centre-of-mass distance gained at 5 ms, after which the intensity of coalescence reduces and viscous dissipation becomes important (see main text). A comparison of the apparent and real roaming velocities for the same set of roaming events is shown in **Figure S7.2a**. Next, we divide the real roaming velocities by the theoretical maximum, which is as if all excess liquid-vapour interfacial energy were converted to in-plane translational kinetic energy of the main droplet. The result is a dimensionless real velocity of 0.48 ± 0.14, compared to 0.17 of droplet jumping.[9] See the left plot of **Figure S7.2b**. Finally, we plot the dimensionless real velocity against the total liquid-vapour interfacial surface energy of coalesced participating droplets at 5 ms for each event, in the right plot of **Figure S7.2b**. An increasing trend is observed, suggesting that closely packed small droplets may benefit the energy conversion.

## S8. Heat flux against subcooling on boehmite

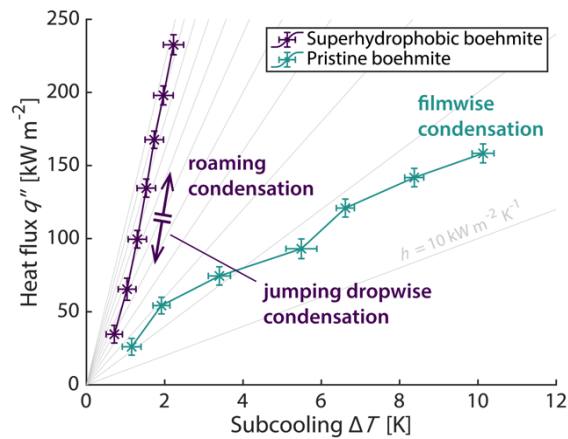

**Figure S8.1**: Measured heat fluxes $q''$ at steady state. Lines of constant heat transfer coefficient $h$ are shown in grey, from 10 to 110 kW m$^{-2}$ K$^{-1}$ at intervals of 10. $dh/d(\Delta T) > 0$ when $dq''/d(\Delta T) > h$. On the superhydrophobic surface, two modes of condensation are observed.

**Figure S8.1** plots the heat flux against subcooling for the superhydrophobic boehmite and pristine boehmite surfaces. The superhydrophobic surface always requires smaller subcooling to sustain the same heat flux. **Figure 3a** plots the heat transfer coefficient against subcooling. On the superhydrophobic surface, roaming condensation (mean = 183.2 kW m$^{-2}$, last 4 points from the left in **Figure S8.1**) sustains a 175% higher heat flux than jumping dropwise condensation (mean = 66.5 kW m$^{-2}$, first 3 points from the left in **Figure S8.1**), owing to the combined effect of a higher thermal driving force, i.e. subcooling, and a higher heat transfer coefficient.

## S9. Rates of condensate volume removal and surface area renewal

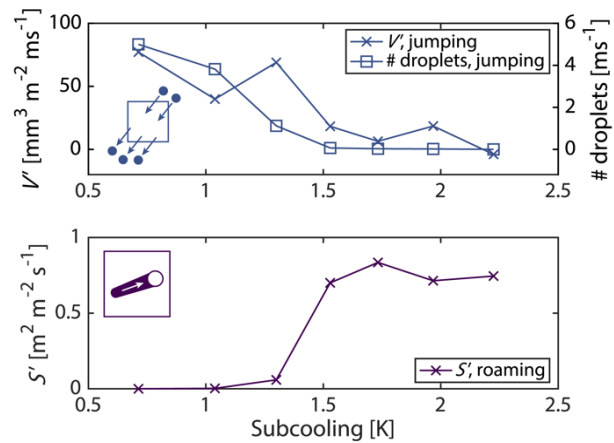

**Figure S9.1**: Condensate volume removal rate $V'$ from jumping, calculated from the difference between the volume of jumped droplets entering and exiting the crop region. Unit: mm³ of condensate per m² of condensing surface per ms of time (top, left axis). Total number of jumped droplets entering and exiting the crop region. Unit: number per ms (top, right axis). Surface area renewal rate $S'$ from roaming. Unit: m² of surface area renewed per m² of condensing surface per s of time (bottom). At ≈ 1.5 K, a transition is seen from the reduction of $V'$ and number of droplets from jumping; and the increase in $S'$ from roaming.

Condensation exhibits a transition as subcooling increases. Refer to **Figure S9.1**. At the transition subcooling of 1.5 K, the condensate volume removal rate $V'$ and the number of droplets from jumping decreases and approaches zero, whereas the surface area renewal rate $S'$ from roaming increases sharply from zero. Measurement procedures are detailed in the following.

To avoid edge effects, we choose to consider a square crop region of size 2250 μm × 2250 μm at the centre of our field of view with a size approximately of 3160 μm × 3160 μm. This field of view is approximately at the centre of the condensing surface of size 20000 μm × 20000 μm. This section provides the details for the measurement of condensate volume removal rate and the number of droplets from jumping, and the surface area renewal rates from roaming.

Condensate volume removal rate and the number of droplets from jumping

The condensate volume removal rate from jumping is measured at different subcooling with ImageJ. As droplets jump from the condensing surface, they are affected by the steam flow and gravity. Therefore, numerous droplets that have jumped away from the surface travel across the field of view and the crop region. To identify the volume of the droplets that jumped from the surface within the crop region, we measure the total volume of jumped droplets entering the crop region (inflow) and the total volume of jumped droplets exiting the crop region (outflow). The difference between outflow and inflow is the condensate volume removal rate from jumping in the crop region.

For the lowest 3 subcooling, 50 ms of video is measured as jumping is the dominant mode. For the highest 4 subcooling, 1000 ms of video is measured as jumping vanishes. When a jumped droplet crosses the boundary of the crop region, it is measured as a circle with the oval tool of ImageJ. The measured diameter is converted to the volume of the jumped droplet as $(4/3)\pi(d/2)^3$. The total volumetric inflow and outflow, and their difference, in the timeframe considered can then be obtained. In **Figure S9.1**, we report the normalised condensate volume removal rate from jumping per area of condensing surface per time. Negative values are possible when in the timeframe considered, more droplets land and come to rest in the crop region than those which jump from the condensing surface. We also report in **Figure S9.1** the total number of jumped droplets entering and exiting the crop region.

The above volumetric method is adopted because jumping is dominant when subcooling is low. Droplets on the surface are small and the surface area renewed from jumping cannot be easily discerned. The condensate volume removal rate provides an estimate for the renewal of the surface. Lastly, it is noted that roaming can terminate in jumping as well around the transition subcooling. The reported removal rate includes all jumping droplets.

Surface area renewal rate from roaming

The surface area renewal rate from roaming is measured at different subcooling with ImageJ.

Roaming motion exhibits clear directionality compared to local multi-droplet coalescence. For each subcooling, 11546 frames are examined for roaming motion events (recorded at 10000 fps, thus duration approximately 1.15 s). At approximately the end of each event (i.e. the main droplet has either stopped or jumped), the polygon tool in ImageJ is used to enclose and measure the renewed surface area, indicated by the lack of condensate droplets and higher reflectivity. Only the renewed surface area within the crop region is measured. A roaming event is only considered when it begins within the field of view, at least one participating droplet is in the crop region, and not initiated by a landing droplet visibly transferring momentum. In **Figure 3c** and **S9.1**, we report the normalised surface area renewal rate from roaming per area of condensing surface per time.

## S10. Critical nucleation diameter and transition subcooling

From classical nucleation theory, the diameter at which nucleation occurs, i.e. the critical nucleation diameter $d_{crit}$, follows the expression:[10]

$$d_{crit} = \frac{4\sigma}{(RT_{surf}/v_l)\ln[P_v/P_{sat}(T_{surf})] - P_v + P_{sat}(T_{surf})}$$

where $\sigma$ is the surface tension, $R$ is the ideal gas constant, $v_l$ is the specific volume of the condensate, $P_v$ is the steam pressure and $P_{sat}(T_{surf})$ is the saturation pressure at the surface temperature. **Figure 3c** plots the variation of the critical nucleation diameter $d_{crit}$ with subcooling for 30 mbar saturated steam, as well as the cavity size distribution of boehmite.

Transition subcooling

Condensation transitions from the jumping dropwise mode to the roaming mode upon increase in subcooling past the transition threshold. This is observed on boehmite (main text) and other types of nanostructures (**Supplementary Information S5**). On titanium dioxide nanorods, the transition subcooling is similar to boehmite ($\approx 1.5$ K), due to a similar density of nanostructures. However, when the nanostructures are sparser resulting in larger cavities on copper(II) hydroxide nanoneedles, the transition subcooling is notably reduced to 0.7 K.

**Figure S10.1** compares the structures of boehmite and copper(II) hydroxide. The density of nanostructures on copper(II) hydroxide is much lower with significantly larger cavities.

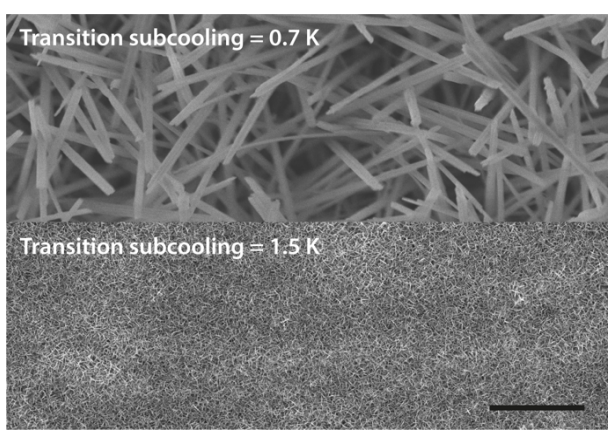

**Figure S10.1**: SEM images of copper(II) hydroxide nanoneedles (top) and boehmite nanowalls (middle), at the same magnification. Both are coated with pPFDA. Scale bar: 2 μm.

As we increase the subcooling on superhydrophobic copper(II) hydroxide nanoneedles, a transition from jumping dropwise condensation to roaming condensation is observed. These structures flood as we further increase the subcooling. **Figure S10.2** shows the heat fluxes $q''$ and heat transfer coefficients $h$ at different subcooling.

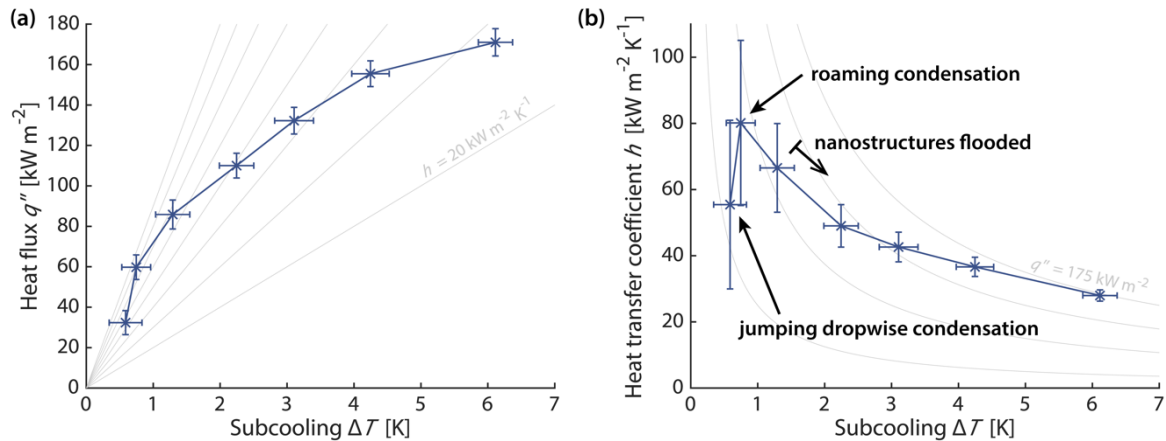

**Figure S10.2**: Heat transfer at different subcooling on copper(II) hydroxide nanoneedles. (a) Measured heat fluxes $q''$ at steady state. Lines of constant heat transfer coefficient $h$ are shown in grey, from 20 to 90 kW m$^{-2}$ K$^{-1}$ at intervals of 10. (b) Heat transfer coefficients $h$ corresponding to **a**. Lines of constant heat flux $q''$ are shown in grey, from 25 to 175 kW m$^{-2}$ at intervals of 50.

At the lowest subcooling measured (0.6 K), condensation is predominantly jumping dropwise. At a higher subcooling of 0.7 K, the condensation mode quickly transitions to roaming. At this point, we measure the highest heat transfer coefficient. However, as subcooling is further increased, the surface floods, accompanied by a gradual decrease in $h$. **Figure S10.3** shows the condensation behaviour at different subcooling.

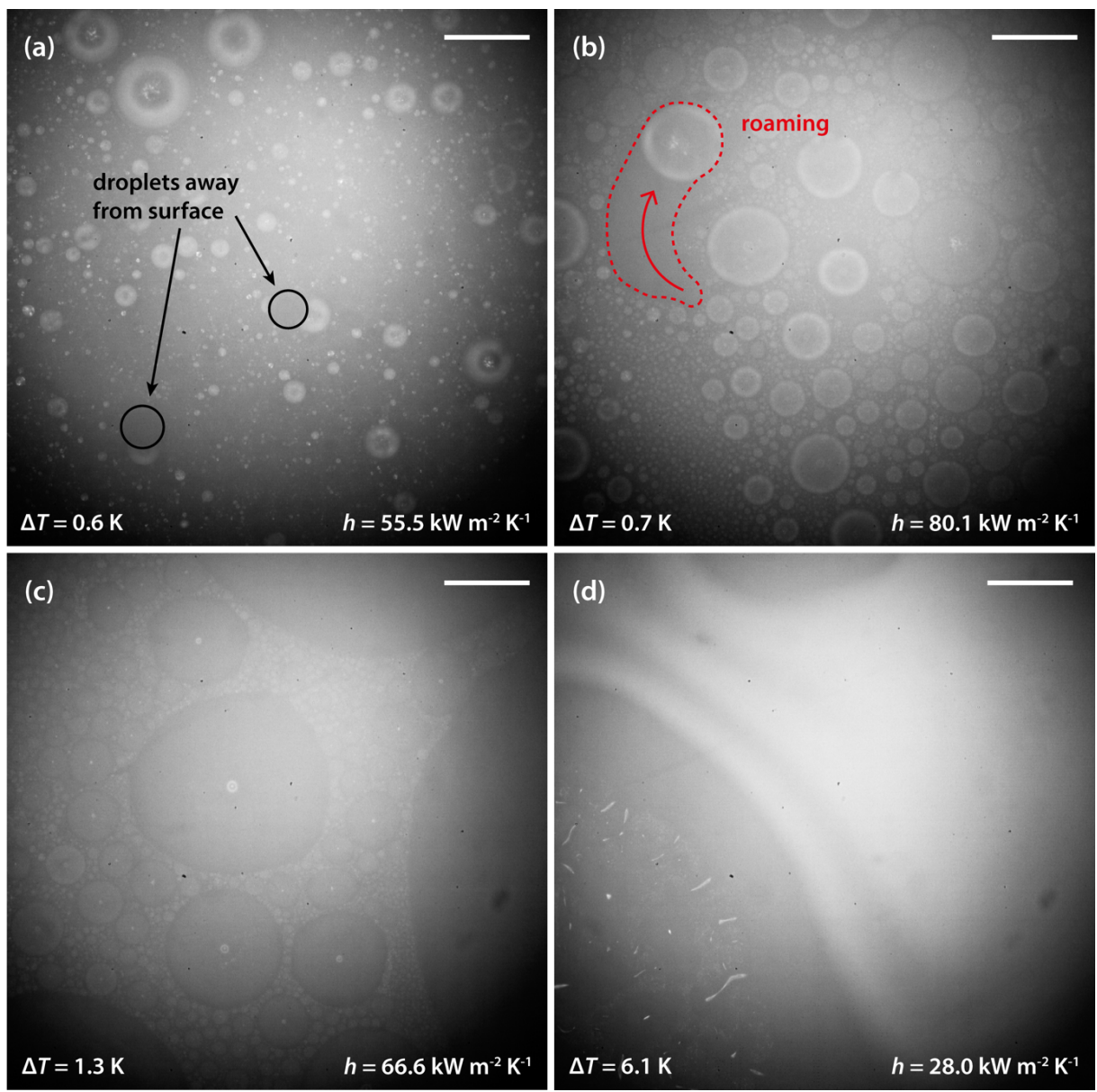

**Figure S10.3**: Snapshots of condensation behaviour at different subcooling on copper(II) hydroxide nanoneedles. Transition is seen from (a) jumping dropwise to (b) roaming condensation, and finally to (c) and (d) flooded condensation. Scale bars: 500 μm.

## S11. Volumetric nucleation rate and cavity filling timescale

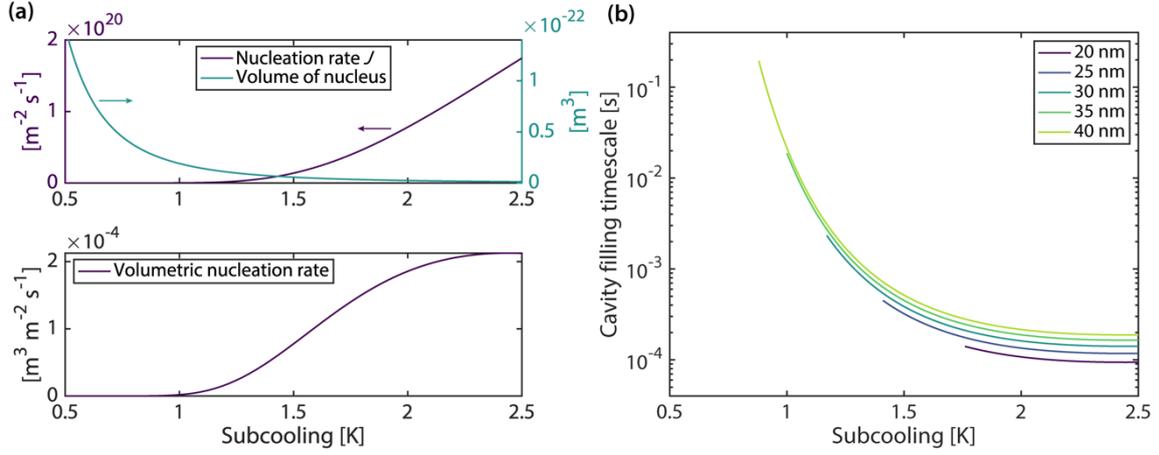

**Figure S11.1**: (a) Nucleation rate $J$ and volume of each nucleus $V_{\text{nucl}}$ (top), and their product, the volumetric nucleation rate (bottom). Transition can be seen at $\approx 1.5$ K. (b) Timescale to fill the nanostructure cavities of sizes 20 – 40 nm. For each size, there is a minimum subcooling to nucleate within it, determined by the critical nucleation diameter.

<u>Volumetric nucleation rate</u>

We extend our model to include the dynamics of nucleation in **Figure S11.1**. We first calculate the nucleation rate $J$ (in number of nuclei per area per time) with the following expressions:[10]

$$J = A \exp\left(\frac{-16\pi(\sigma F_{\text{eff}}/k_B T_{\text{surf}})^3 (\bar{M} v_l/N_A)^2}{3\{\ln[P_v/P_{\text{sat}}(T_{\text{surf}})]\}^2}\right)$$

where

$$A = \left(\frac{2\sigma F_{\text{eff}} N_A}{\pi \bar{M}}\right)^{1/2} \left(\frac{P_v}{RT_{\text{surf}}}\right)^{5/3} \left(\frac{N_A}{\bar{M}}\right)^{2/3} v_l F_{\text{eff}} \left(\frac{1 - \cos\theta_{\text{eff}}}{2}\right)$$

and

$$F_{\text{eff}} = \frac{2 - 3\cos\theta_{\text{eff}} + (\cos\theta_{\text{eff}})^3}{4}$$

Here, $\theta_{\text{eff}}$ is the effective contact angle. $\theta_{\text{eff}}$ is expected to be much smaller than the intrinsic contact angle of pPFDA ($\theta_{\text{int}} = 119.2°$, see **Methods**) because the energetic barrier to nucleation within cavities is significantly lower than that on a flat surface.[11] In the expression for $J$, $N_A$ is the Avogadro's number, $\bar{M}$ is the molecular weight, and $k_B$ is the Boltzmann constant.

The volumetric nucleation rate, i.e. the total volume of nuclei per area per time, is obtained by multiplying the nucleation rate $J$ with the volume of each nucleus. We calculate the volume of each nucleus $V_{\text{nucl}}$ at the critical nucleation diameter $d_{\text{crit}}$ and assuming a spherical cap shape with the surface at the intrinsic water-pPFDA contact angle $\theta_{\text{int}} = 119.2°$ using the following expression:

$$V_{\text{nucl}} = \frac{\pi}{3} \left(\frac{d_{\text{crit}}}{2}\right)^3 (2 + \cos\theta_{\text{int}})(1 - \cos\theta_{\text{int}})^2$$

We find that for an effective contact angle $\theta_{\text{eff}} = 48°$, a transition in the volumetric nucleation rate is seen around a subcooling of 1.5 K (**Figure S11.1a**).

Cavity filling timescale

The structure of boehmite consists of cavities sided by nanowalls (right inset in **Figure 1a**). The time for nucleation to fill the cavity $\tau_{\text{fill}}$ can therefore be calculated as follows:

$$\tau_{\text{fill}} = \frac{V_{\text{cav}}}{JA_{\text{cav}}V_{\text{nucl}}}$$

where $V_{\text{cav}}$ and $A_{\text{cav}}$ are the volume and surface area of a cavity respectively. As $V_{\text{cav}}$ and $A_{\text{cav}}$ scale with the characteristic length scale of the cavity $L_{\text{cav}}$ as $V_{\text{cav}} \sim L_{\text{cav}}^3$ and $A_{\text{cav}} \sim L_{\text{cav}}^2$, we then calculate the cavity filling timescale $\tau_{\text{fill}}$ as a function of $L_{\text{cav}}$ as follows:

$$\tau_{\text{fill}} \sim \frac{L_{\text{cav}}^3}{JL_{\text{cav}}^2 V_{\text{nucl}}} = \frac{L_{\text{cav}}}{JV_{\text{nucl}}}$$

We compute the range of $L_{\text{cav}}$ for boehmite by measuring the projected area of each cavity in an SEM image using the polygon tool in ImageJ. $L_{\text{cav}}$ is then taken as the square root of the measured projected area. The distribution of $L_{\text{cav}}$ for boehmite is shown in **Figure 3c**.

Substituting the respective equations, the timescale to fill the nanoscale cavities of boehmite is estimated to be 0.1 – 1 ms (**Figure S11.1b**).

## S12. Effects from droplet size mismatch

A smaller droplet has a higher Laplace pressure and a lower mass, enabling it to complete its coalescence in a shorter time compared to the larger droplet, and resulting in motion which appears to be tangential as the smaller droplet is apparently coalesced "into" the larger one. However, if the two droplets are considered as a single system, without roaming, there is little net tangential movement of its centre of mass after coalescence, as we find in **Figure S7.1**. In this section, we show that tangential momentum generation is minimal from the coalescence of size-mismatched droplets.

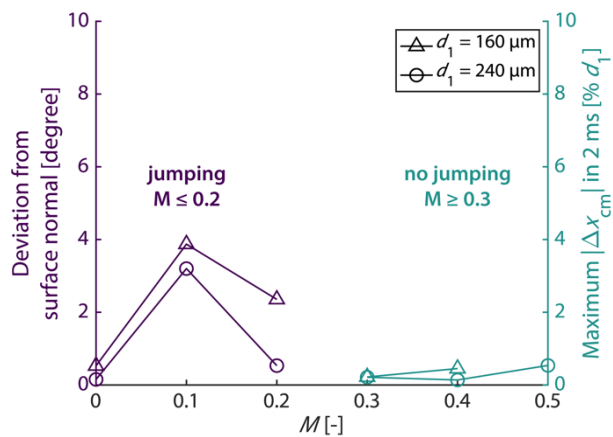

**Figure S12.1**: Coalescence of size-mismatched droplets. Left y-axis: Jumping occurs when mismatch $M \leq 0.2$. The deviation from surface normal of the jumping motion is shown. Minimal deviation (< 4°) is observed. Right y-axis: No jumping occurs when $M \geq 0.3$. The maximum absolute centre-of-mass displacement in the in-plane x-direction in the 2 ms simulated is shown, normalised by the diameter of the smaller droplet $d_1$. Minimum displacement (< 1%) is observed.

Through numerical simulations in **Figure S12.1** we predict the direction of motion of the liquid body after the coalescence of two droplets of different mismatch ratios, defined as $M = (d_2 - d_1)/(d_1 + d_2)$, where $d_1$ and $d_2$ are the diameters of the smaller and larger droplets respectively. Mismatch ratios from 0 to 0.5 are tested for $d_1$ = 240 µm and 0 to 0.4 are tested for $d_1$ = 160 µm. For both $d_1$, no jumping departure from the surface is observed for mismatch ratios of 0.3 or above.

For cases in which there is jumping departure ($M \leq 0.2$), we report the deviation from surface normal of the jumping motion. Our cases tested record deviations smaller than 4° (left y-axis of **Figure S12.1**), in line with a previous experimental study on CuO nanostructured superhydrophobic surfaces[12] which reported a maximum deviation of 3.8°. For cases in which there is no jumping departure ($M \geq 0.3$), we report the maximum absolute centre-of-mass displacement in the in-plane x-direction in the 2 ms simulated, normalised by $d_1$ of the respective case. Our cases tested record maximum normalised displacements smaller than 1% (right y-axis of **Figure S12.1**). It is therefore clear that substantial tangential momentum cannot be generated from size mismatch at the droplet level.

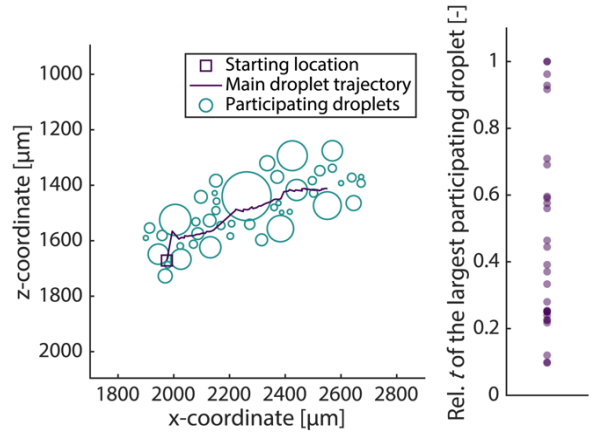

**Figure S12.2**: Left: participating droplets distribution of a roaming event. Right: the relative time of the largest participating droplet in each measured roaming event.

At the event level, there is as well no observable trend in participating droplet sizes. The largest participating droplet of a roaming event can be at any location along its trajectory. The left panel in **Figure S12.2** illustrates the distribution of participating droplets of a roaming event. For this particular event, the largest participating droplet is at the middle of the trajectory. The right panel plots the distribution of the time of the largest participating droplet in each measured roaming event, relative to the duration of the event. For example, for a relative time of 1, the coalescence of the largest participating droplet occurs at the end (in terms of time) of the roaming event. The scattered distribution found in the right panel indicates that a roaming event is not driven by a trend in participating droplet sizes. Therefore, substantial tangential momentum is not generated by droplet size mismatch at the event level as well.

## S13. Stages of x-momentum generation

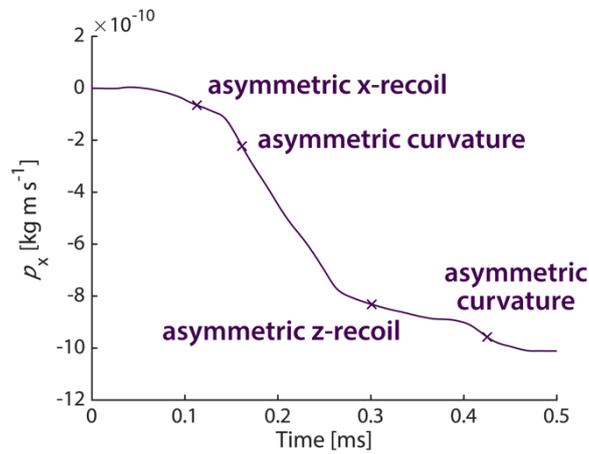

**Figure S13.1**: x-momentum for the first 0.5 ms. Stages of momentum generation are seen. See also **Figure 4b** and **Video S6**.

We plot the variation of x-momentum for the first 0.5 ms in **Figure S13.1**. Tangential momentum generation roughly follows two rates. Asymmetric recoil, originating from wettability difference, generates less momentum compared to asymmetric curvature, which originates from the elongation in the y-direction due to symmetry breaking by the substrate.

## S14. Dewetting and the efficiency in kinetic energy conversion

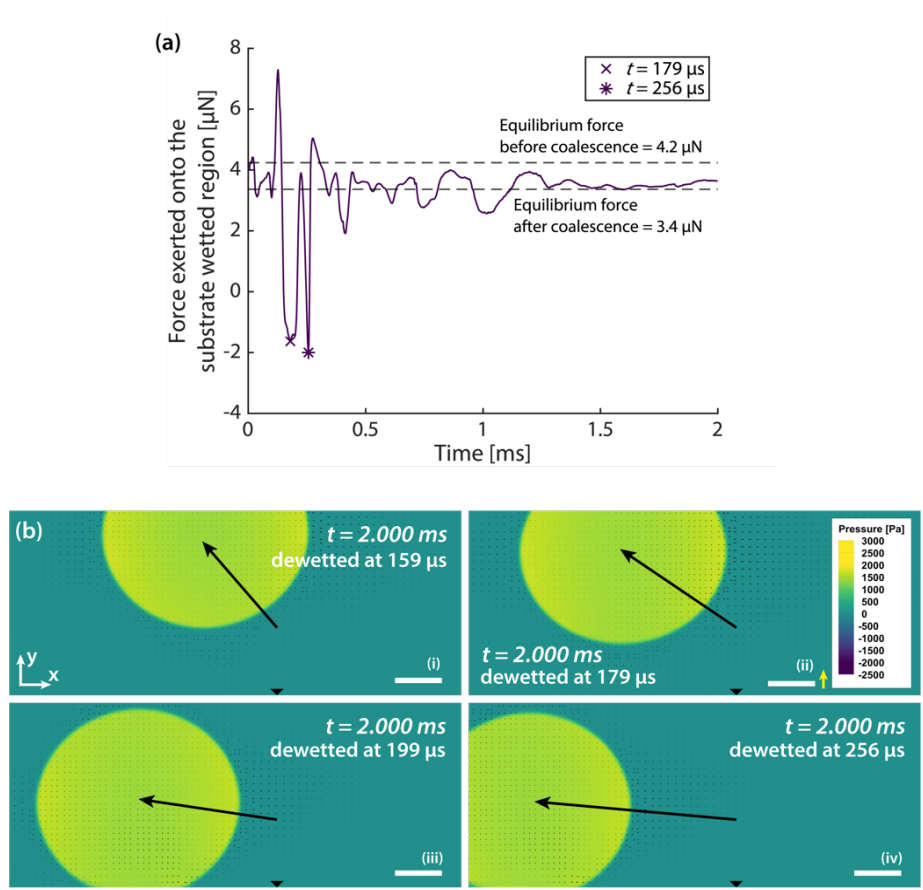

**Figure S14.1**: (a) Force exerted onto the wetted area when there is no dewetting throughout the 2 ms (case in **Figure 4b**). Due to Laplace pressure, the equilibrium forces are non-zero, i.e. the liquid body exerts a force onto the wetted area from the curvature of the interface despite the lack of gravity. The force reported reflects a full circular wetted region, taking domain symmetry into account. (b) Droplet location at 2 ms for different dewetting times. All departures contain a substantial tangential component. Black arrows indicate the direction of motion. Markers carry the same meaning as in **Figure 4b**.

When two droplets coalesce and the base area of one of them is wetted, a pulling force is exerted onto the wetted region at some point during the coalescence. **Figure S14.1a** displays the force on this region when there is no dewetting throughout 2 ms (case in **Figure 4b**).

In the beginning, a force of ≈ 4.2 µN is exerted. It is due to the Laplace pressure of a 160 µm diameter droplet on the wetted region:

$$\text{Force} = (2\sigma/r_{\text{curv}})A_{\text{wetted}}$$

where $\sigma = 0.072$ N m$^{-1}$, $r_{\text{curv}} = 160/2 = 80$ µm is the radius of curvature and $A_{\text{wetted}} = \pi \left(\frac{160}{2}\sin 160°\right)^2$ is the area of the wetted region.

Then, as the two droplets coalesce, a pulling force is exerted onto the wetted region. There are two peaks, as indicated by a cross and an asterisk in **Figure S14.1a**. In **Figure 5**, we choose to dewet at 179 µs and substantial x-displacement in the departure is seen.

As there is no dewetting for the case corresponding to the plot in **Figure S14.1a**, the two droplets complete coalescence and come to rest on top of the wetted region. At this point it exerts a lower force than in the beginning, as the new radius of curvature ($r_{\text{curv}} = 101$ µm) is larger with the combined volume of the two droplets. Using the equation above, it can be calculated that the equilibrium force after coalescence with the new radius of curvature is 3.4 µN. In **Figure S14.1a** we can observe that the force from our simulation converges to this value.

**Figure S14.1b** displays the location of the droplet at 2 ms for different dewetting times. We choose to dewet at the two peaks in **Figure S14.1a**, as well as two additional times at 159 and 199 µs. It can be seen that the later is the dewetting, the higher the x-displacement is. **Figure S14.2** displays the kinetic energy of the translational motion of the centre of mass compared to the total kinetic energy, for different dewetting times and when there is no dewetting. Similarly, the later is the dewetting, the closer $KE_{\text{cm}}$ is to $KE_{\text{tot}}$, indicating a higher energy conversion efficiency. This arises from the fact that for wetting cases, the y-centre of mass is consistently lower, and the liquid body is closer to the surface. As it oscillates, there is stronger symmetry

breaking than the reference case where the droplet is departing normally from the surface earlier in the coalescence process.

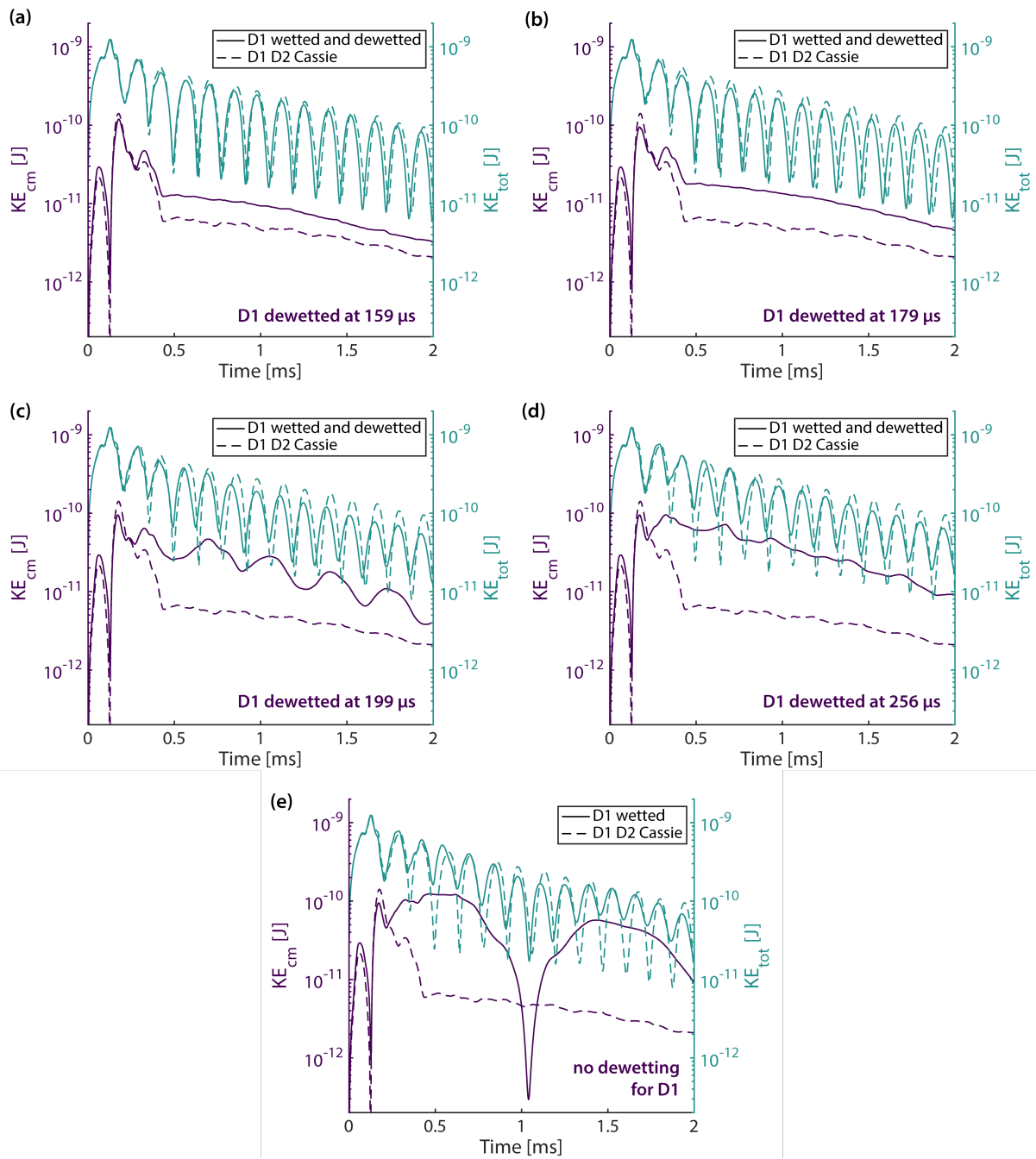

**Figure S14.2**: Kinetic energy of the translational motion of the centre of mass $KE_{cm}$ and the total kinetic energy $KE_{tot}$ for (a) to (d) different dewetting times, and (e) no dewetting. The kinetic energy reported reflects full spherical droplets, taking domain symmetry into account.

**S15. Setup of numerical simulation cases**

Numerical studies are performed with Ansys Fluent 2021 R2 on the Euler computing cluster of ETH Zurich. This section outlines the setup of simulation cases involving two droplets, with or without droplet size mismatch.

Domain and meshing

The three-dimensional computational domain is created and meshed with hexahedra in Ansys ICEM CFD 2021 R2. An example can be seen in **Figure 4a**.

The domain is in the shape of a cuboid and split along a symmetry plane, where the centre-to-centre line of the two droplets lie. Nanostructures cannot be explicitly included in the geometry, but modelled as a flat superhydrophobic surface on the x-z plane. The symmetry plane is normal to the z-axis.

The domain size is specified as multiples of the mean droplet radius. The x-, y- and z-dimensions are $(12,10,6)$ $\bar{r}$ respectively,[13] where $\bar{r}$ is the mean radius of the two droplets. Cells are cubes with equal edge lengths $\Delta h$ in all dimensions. The maximum global cell size is specified as 10 cells per radius of the smaller droplet. For cells around the liquid-vapour interface and within the droplet(s), their size is refined to 40 cells per radius of the smaller droplet,[13] through solution-based adaptation described later.

Model and fluid properties

The multiphase flow is modelled with the volume of fluid (VOF) method. Two discrete phases are specified, namely water vapour and liquid water. The density and viscosity for water vapour are 0.021904 kg m$^{-3}$ and 9.6718 × 10$^{-6}$ kg m$^{-1}$ s$^{-1}$, and the density and viscosity for liquid water are 997.24 kg m$^{-3}$ and 9.0904 × 10$^{-4}$ kg m$^{-1}$ s$^{-1}$, corresponding to their saturation properties at 30 mbar, the pressure maintained in the experiments. These properties are obtained with CoolProp.[14] The surface tension is taken to be 0.072114 N m$^{-1}$, provided by The International Association for the Properties of Water and Steam,[15] and similarly at saturation at 30 mbar. The continuum surface force model[16] is used to model surface tension and gravity is disabled.

Boundary conditions

The superhydrophobic surface is modelled as a no-slip wall with a specified contact angle. Symmetry is specified at the symmetry plane of the droplets. The three vertical boundaries are specified as pressure inlets with a gauge pressure of zero and normal flow direction. Similarly, the top horizontal boundary is specified as a pressure outlet with a gauge pressure of zero. Gauge pressures are computed using an operating pressure of 30 mbar.

Solver

A double-precision pressure-based transient solver is used. The PISO scheme is specified for pressure-velocity coupling. For spatial discretisation, gradients are computed using the least squares cell based method. The scheme for pressure is PRESTO!, second order upwind for momentum, and geo-reconstruct for volume fraction. For temporal discretisation, the first order implicit scheme is used. To determine convergence, the residuals of continuity and all three velocities should reach 10$^{-6}$ to proceed to the next time step. The total mass balance across all flow boundaries is monitored as well.

Mesh adaptation

A mesh adaptation strategy is adopted to refine the mesh around and within the droplet(s) to resolve and capture dynamics at small scales for the droplet(s). A cell marked for refinement satisfies at least one of the two criteria:

1) The gradient of the liquid water volume fraction is larger than 6% of the global maximum, so that the cell is refined if it is close to any liquid/vapour interface; or

2) The liquid water volume fraction is larger than the cut-off value of $10^{-6}$, so that the cell is refined if it is occupied with liquid water.

Cells marked for refinement are refined to 1/4 of their original edge length, corresponding to 1/64 of their original cell volume. As the solution proceeds and a cell becomes occupied with only water vapour and far away from any liquid/vapour interface, it is coarsened back to its original edge length. A cell marked for coarsening has to satisfy both criteria:

1) The gradient of the liquid water volume fraction is smaller than 5% of the global maximum, so that the cell is away from any liquid/vapour interface; and

2) The liquid water volume fraction is smaller than the cut-off value of $10^{-6}$, so that the cell is occupied with water vapour only.

The mesh is pre-adapted during solution initialisation using the criteria above. During calculation, the adaptation is updated every two time steps based on the instantaneous solution flow field, refining and coarsening the concerned cells.

Solution initialisation

The entire flow field is first initialised with zero gauge pressure, velocity and liquid water volume fraction. Then, cells which constitute a droplet are patched with a liquid water volume fraction of one. The mesh is adapted repeatedly with the criteria above until no more cells are marked for refinement or coarsening. However, note that this resulting mesh is adapted from a liquid droplet patched onto the initial coarse mesh, i.e. the droplet is approximated by a coarse interface to start with. As mesh adaptation does not alter the flow field, the droplet remains coarse albeit on the refined, adapted mesh. If time marching begins here, it would take a few time steps for the coarse droplet and its liquid-vapour interface to relax into the refined mesh, distorting the solution. Therefore, to ensure that the droplet is patched onto a sufficiently refined mesh so that its contour is well resolved in the beginning, we repeat the initialisation, patching and adaptation procedure until the mesh adaptation criteria no longer mark any cells for refinement or coarsening after patching the liquid droplet cells.

Solution computation and time advancement

With the domain and solution properly initialised, the case can proceed to computation. To capture all dynamics, we select a time step size smaller than the timescale of relevant physical phenomena, derived at the length scale of the refined cell size. Among surface tension, convective flow, and viscous dissipation, the timescale for surface tension is found to be the smallest, computed as follows:[16]

$$t_\sigma = \sqrt{\frac{\bar{\rho}(\Delta h)^3}{2\pi\sigma}}$$

where $\bar{\rho}$ is the mean density of the two phases, $\Delta h$ is the cell edge length, and $\sigma$ is the surface tension. For a case in which the smaller droplet has a radius of 80 μm, the adapted cell edge length is 2 μm and the resulting surface tension timescale is 94 ns. A time step size of 90 ns is then specified for the computation.

Result reporting and postprocessing

During computation, results including droplet centre of mass, momentum and kinetic energy are reported and saved every time step. Selected variables of the flow field solution is saved at least every 1 μs and the entire flow field solution is saved at least every 10 μs. Postprocessing is performed in Tecplot 360 EX 2021 R2 and MATLAB R2022b (MathWorks).

Computation of variables of interest

Several variables are computed during the simulation and in postprocessing. We report the expression for the computation of a number of them in the following.

1) Centre-of-mass location

The x-coordinate of the centre of mass of the liquid body is calculated as follows:

$$x_{cm} = \frac{\int_\Omega \rho \alpha x \, d\Omega}{\int_\Omega \rho \alpha \, d\Omega}$$

where $d\Omega$ is the volume of a cell in the domain, $\rho$ is the density of the liquid, $x$ is the x-coordinate of the cell and $\alpha$ is the volume fraction of the liquid of the cell. The integrals integrate over the entire computational domain $\Omega$. The y- and z-coordinates of the centre of mass of the liquid body are calculated similarly, replacing $x$ with $y$ and $z$ respectively.

2) Momentum

The total x-momentum of the liquid body is calculated as follows:

$$p_x = \int_\Omega v_x \rho \alpha \, d\Omega$$

Where $v_x$ is the velocity in the x-direction of the cell. The total y- and z-momentum of the liquid body are calculated similarly, replacing $v_x$ with $v_y$ and $v_z$ respectively.

3) Total kinetic energy

The total x-kinetic energy of the liquid body is calculated as follows:

$$KE_{x,\,tot} = \int_\Omega \frac{1}{2} \rho \alpha v_x^2 \, d\Omega$$

The total y- and z-kinetic energies of the liquid body are calculated similarly, replacing $v_x$ with $v_y$ and $v_z$ respectively. The total kinetic energy of the liquid body is the sum of all three directions.

4) Kinetic energy of the translational motion of the centre of mass

The x-kinetic energy of the translational motion of the centre of mass of the liquid body is calculated as follows:

$$KE_{x,\,cm} = \frac{1}{2} \left( \int_\Omega \rho \alpha \, d\Omega \right) \left( \frac{dx_{cm}}{dt} \right)^2$$

The integral is the total mass of the liquid body and $dx_{cm}/dt$ is the velocity of the centre of mass of the liquid body in the x-direction. The y- and z-kinetic energies of the translational motion of the centre of mass of the liquid body are calculated similarly, replacing $x_{cm}$ with $y_{cm}$ and $z_{cm}$ respectively. The kinetic energy of the translational motion of the centre of mass is the sum of all three directions.